\documentclass[11pt]{article}
\usepackage{amsmath}

\usepackage{rotating}
\usepackage[small]{caption}
\usepackage[numbers,square]{natbib}

\usepackage{graphics}
\usepackage{amssymb}
\usepackage{subfigure}
\usepackage{epsf}
\usepackage{rotating}
\usepackage{pstricks}
\def\ba{\begin{eqnarray}}
\def\ea{\end{eqnarray}}

 \let\a=\alpha      \let\g=\gamma

 \let\n=\nu

\def\slashed{\ds}

\def\a{\alpha}

\def\g{\gamma}

\def\ds#1{#1\kern-1ex\hbox{/}}
\def\dsh{h\kern-1.2ex /}

\let\G=\Gamma

\newcommand{\bea}{\begin{eqnarray}}
\newcommand{\eea}{\end{eqnarray}}

\def\nn{\nonumber}
\def\beq{\begin{equation}}
\def\eeq{\end{equation}}

\def\beqn{\begin{eqnarray}}
\def\eeqn{\end{eqnarray}}
\def\ba{\begin{eqnarray}}
\def\ea{\end{eqnarray}}

\def\slash#1{#1\hskip-6pt/\hskip6pt}

\newcommand{\beqa}{\begin{eqnarray}}
\newcommand{\eeqa}{\end{eqnarray}}

\begin{document}

\begin{center}
\vspace{1.5cm}

{\bf\Large{Dark Matter with St\"uckelberg Axions } }

\vspace{0.5cm}
\vspace{1cm}
{\bf $^{(1)}$Claudio Corian\`o, $^{(1)}$Paul H. Frampton, $^{(2)}$Nikos Irges and $^{(1)}$Alessandro Tatullo\\}
{\it  $^{(1)}$Dipartimento di Fisica, Universit\`{a} del Salento \\
and  INFN Sezione di Lecce,  Via Arnesano 73100 Lecce, Italy\\}
{\it $^{(2)}$Department of Physics\\
National Technical University of Athens\\
Zografou Campus, GR-15780 Athens, Greece}

\begin{abstract}
We review a class of models which generalize the traditional Peccei-Quinn (PQ) axion solution by a St\"uckelberg pseudoscalar. 
Such axion models represent a significant variant with respect to earlier scenarios where axion fields were associated with global anomalies, because of the St\"uckelberg 
field, which is essential for the cancellation of gauge anomalies in the presence of extra $U(1)$ symmetries. The extra neutral currents associated to these models have been 
investigated in the past in orientifold models with intersecting branes, under the assumption that the St\"uckelberg scale was in the multi-TeV region. Such constructions, 
at the field theory level, are quite general and can be interpreted as the four-dimensional field theory realization of the Green-Schwarz mechanism of anomaly cancellation of string theory. 
We present an overview of models of this type in the TeV/multi TeV range in their original formulation and their recent embeddings into an ordinary GUT theory, presenting an 
$E_6\times U(1)_X$ model as an example. In this case the model contains two axions, the first corresponding to a Peccei-Quinn axion, whose misalignment takes place 
at the QCD phase transition, with a mass in the meV region and which solves the strong CP problem. The second axion is ultralight, in the $10^{-20}-10^{-22}$ eV region, 
due to a misalignment and a decoupling taking place at the GUT scale. The two scales introduced by the PQ solution, the PQ breaking scale and the misalignment scale at the 
QCD hadron transition, become the Planck and the GUT scales respectively, with a global anomaly replaced by a gauge anomaly.  The periodic potential and the corresponding oscillations are related to a particle whose
De Broglie wavelength can reach 10 kpc. Such a sub-galactic scale has been deemed necessary in order to resolve several dark matter issues at the astrophysical level. 

\end{abstract}
\end{center}

\newpage
\section{Introduction}

 It is by now well established that astrophysical and cosmological data coming either from measurements of the velocities of stars orbiting galaxies, in their rotation curves, or from the cosmic microwave background, 
 indicate that about $\sim 80 \%$ of matter in the universe is in a unknown form, and the expectations for providing an answer to such a pressing question run high. These observational results are justified within the 
 standard $\Lambda$CDM dark matter/dark energy model \cite{Ade:2015xua} which has been very successful in explaining the data. It predicts a dark energy component about $68 \pm 1 \%$ of the total mass/density 
 contributions of our universe in the form of a cosmological constant. The latter accounts for the dark energy dominance in the cosmological expansion at late times and provides the cosmological acceleration measured by 
 Type Ia supernovae \cite{Riess:1998cb,Perlmutter:1998np}, with ordinary baryonic dark matter contributing just a few percent of the total mass/energy content ($\sim 5 \%$) and a smaller neutrino component. 
 Cold dark matter with small density fluctuations, growing gravitationally and a spectral index of the perturbations $n_S\sim 1$ is compatible with an early inflationary stage and accounts for structure formation in most of 
 the early universe eras. By now, data on the CMB, weak lensing and structure formation, covering redshifts from large $z\sim10^3$ down to $z <\sim O(1)$ where the full nonlinear regime of matter dominance is at work, 
 have been confronted with N-body gravitational simulations for quite some time, with comparisons which are in general agreement with $\Lambda$CDM.
Such simulations, characterized by perturbations with the above value of the spectral index show the emergence of hierarchical, self-similar structures in the form of halos and sub-halos of singular density $(\rho(r)\sim 1/r$ 
in terms of the radius $r$) \cite{Navarro:1996gj} in the nonlinear regime. However, while the agreement between $\Lambda$CDM and the observations is significant at most scales, at a small sub-galactic scale, 
corresponding to astrophysical distances relevant for the description of the stellar distributions ($\sim 10$ kpc), cold dark matter models predict an abundance of low-mass halos in excess of observations \cite{Hu:2000ke}. 
Difficulties in characterizing this sub-galactic region have usually been attributed to inaccurate modeling of its baryonic content, connected with star formation, supernova explosions and black hole activity which take place in that region, causing a redistribution of matter.

There are various possibilities to solve this discrepancy, such as invoking the presence of warm dark matter (WDM), whose free streaming, especially for low mass WDM particles, could erase halos and sub-halos of low mass. 
At the same time they could remove the predicted dark matter cusps in $\rho(r)$, present in the simulations for $r\simeq 0$ \cite{Navarro:1996gj} but not detected observationally.
As observed in \cite{Hu:2000ke} and recently re-addressed in \cite{Hui:2016ltb}, these issues define a problem whose resolution may require a cold dark matter component which is ultralight, in the $10^{-20}-10^{-22}$ eV range. 
Proposals for such component of dark matter find motivations mostly within string theory, where massless moduli in the form of scalar and pseudoscalar fields abound at low energy. 
They are introduced at the Planck scale and their flat potentials can be lifted by a small amount, giving rise to ultralight particles. However, the characterization of a well-defined gauge structure which may account for the generation 
of such ultralight particle(s) and which may eventually connect the speculative scenarios to the electroweak scale can be pursued in various ways. It has been recently proposed  \cite{Coriano:2017ghp} that particles of this kind may 
emerge from grand unification in the presence of anomalous abelian symmetries, revisiting previous constructions. 

The goal of this review is to summarize the gauge structure of these models which require an anomalous fermion spectrum with gauge invariance restored by a Wess-Zumino interaction, by the inclusion of a St\"uckelberg axion. 
Such models can be thought as the field theory realization of the mechanism of anomaly cancellation derived from string theory. The models reviewed here are characterized by some distinctive key features that we are going to discuss, 
establishing their relation to the Peccei-Quinn model, of which they are an extension at a field theory level.

\section{Anomalous U(1)'s }

The Peccei-Quinn (PQ) mechanism, proposed in the 1970's to solve the strong CP problem \cite{Peccei:1977ur,Sikivie:2006ni,Kim:2008hd} had been originally realized by assigning an 
additional abelian chiral charge to the fermion spectrum of the Standard Model (SM). Alternatively, a similar symmetry can be present in a natural way in specific gauge theories based on groups 
of higher rank with respect to the SM gauge group. This is the case, for instance of the $U(1)_{PQ} $ symmetry found in the $E_6$ GUT discussed in \cite{Frampton:1981ik} (as well as in other realizations), naturally present 
in this theory and which can lead to a solution of the strong CP problem.

As we are going to discuss, the mass of the axion, either in the presence of global or local anomalies is connected to the instanton sector of a non-abelian theory and it is crucial for the mechanism 
of misalignment to be effective that the axion couples to the gauge sector of the same theory.
 In fact, the possibility that more than one axion is part of the spectrum of a certain gauge theory is not excluded, with the mass of each axion controlled by independent mechanism(s) of vacuum 
 misalignment induced at several scales, if distinct gauge couplings for each of such particles with different gauge sectors are present \cite{Coriano:2010py,Coriano:2010ws}.
 We will illustrate this point in the extended $E_6$ theory that we will overview in the next sections, where the inclusion of an extra anomalous $U(1)$ gauge symmetry realizes such a scenario. 
 Different mechanisms of vacuum misalignment may be held responsible for the generation of axions of different masses, whose sizes may vary considerably. 
  
 \subsection{Anomaly cancellation at field theory level with an axion}
 
 In the case of a St\"uckelberg axion, as already mentioned, the PQ symmetry is generalized from global to a local gauge symmetry and the Wess Zumino interactions are needed for the restoration of 
 gauge invariance of the effective action. Such generalizations, originally discussed in the context of low scale orientifold models \cite{Coriano:2005js}, where anomalous abelian symmetries emerge from 
 stacks of intersecting branes, have been proposed in the past as possible scenarios to be investigated at the LHC \cite{Coriano:2006xh,Coriano:2007fw,Coriano:2007xg,Armillis:2008vp,Coriano:2008pg,Coriano:2009zh}, together with their supersymmetric extensions \cite{Coriano:2008aw,Coriano:2008xa,Coriano:2010ws}. 
While anomalous abelian symmetries are interesting in their own right, especially in the search for extra neutral currents at the LHC \cite{Armillis:2007tb,Coriano:2008wf,Armillis:2008vp} \cite{Anastasopoulos:2008jt}, 
one of the most significant aspects of such anomalous extensions is in fact the presence of an axion which is needed in order to restore the gauge invariance of the effective action. 
It was called the "axi-Higgs" in \cite{Coriano:2005js} \cite{Coriano:2006xh} - for being generated by the mechanism of Higgs-St\"uckelberg mixing in the CP-odd scalar sector, induced by a PQ-breaking periodic potential, 
later studied for its implications for dark matter in \cite{Coriano:2010py}. The appearance of such a potential is what allows one component of the St\"uckelberg field to become physical.
 A periodic potential can be quickly recognized as being of instanton origin and related to the $\theta$-vacuum of Yang-Mills theory and can be associated with phase transitions in non-abelian theories. 
 Recent developments have taken into consideration the possibility that the origin of such a potential of this form can be set at a very large scale, such as the scale of grand unification (GUT). Its size is related 
 to the value of the gauge coupling at the GUT scale, characterized by a typical instanton suppression, where the mechanism of vacuum misalignment takes place. 
 
 \subsection{An ultralight axion}
 
In the case of a misalignment generated at the GUT scale, the mass of the corresponding axion is strongly suppressed and can reach the far infrared, in the range of $10^{-20}-10^{-22}$ eV, which is in the optimal range for a possible resolution of several 
astrophysical issues, such as those mentioned in the introduction \cite{Hui:2016ltb}. Proposals for a fuzzy component of dark matter require a weakly interacting particle in that mass range. As in the PQ (invisible axion) case, also in this case two scales are needed in order to realize a similar scenario. 
In the PQ case the two scales correspond to $f_a$, the large PQ breaking scale and the hadronic scale which links the axion mass, $f_a$, the pion $m_\pi$ and the light quarks masses $m_u,m_d$, in an expression that we will summarize below.
In the case of St\"uckelberg axions these fields can be introduced as duals of a 2-form ($B_{\mu\nu}$), defined at the Planck scale ($M_P$) and coupled to the field strength $(F)$ of an anomalous gauge boson via a $B\wedge F$ interaction \cite{Coriano:2017ghp}. \\
The mechanism of Higgs-axion mixing and the generation of the periodic potential can take place at a typical GUT scale. It is precisely the size of the potential at the GUT scale, which is 
controlled by the $\theta$-vacuum of the corresponding GUT symmetry, which is responsible for the generation of an ultralight axion in the spectrum. As already mentioned, in the model discussed 
in \cite{Coriano:2017ghp} a second axion is present, specific to the $E_6$ part of the $E_6\times U(1)_X$ symmetry, which is sensitive to the $SU(3)$ colour sector of the Standard Model after spontaneous symmetry breaking. 
This second field takes the role of an ordinary PQ axion and solves the strong CP problem. 
We will start by recalling the main features of the PQ solution, in particular the emergence of a mass/coupling relation in such a scenario which narrows the window for axion detection down and gets enlarged in the presence 
of a gauge anomaly in St\"uckelberg models \cite{Coriano:2009zh}. We will then turn, in the second part of this review, to a discussion of the St\"uckelberg extension. We will describe the features of such models in their 
non-supersymmetric formulation. Their supersymmetric version requires a separate discussion, for predicting both an axion and a neutralino as possible dark matter relics \cite{Coriano:2008xa,Coriano:2010ws}.  

\section{The invisible PQ axion }

The theoretical prediction for the mass range in which to locate a PQ axion is currently below the eV region. 
The PQ solution to the strong CP problem has been formulated according 
to two main scenarios involving a light pseudoscalar $(a(x))$ which nowadays take the name from the initials of the proponents, the KSVZ axion (or hadronic axion) 
and the DFSZ \cite{Dine:1981rt, Zhitnitsky:1980tq} axion, the latter introduced in a model which requires, in addition, a scalar sector with two Higgs doublets $H_u$ and $H_d$, besides the PQ complex scalar $\Phi$.\\
The small axion mass is attributed to a vacuum misalignment mechanism generated by the structure of the QCD vacuum at the QCD phase transition, which causes a tilt in the otherwise flat PQ potential. 
The latter undergoes a symmetry breaking at a scale $v_{PQ}$, in general assumed to lay above the scales of inflation $H_I$ and of reheating $(T_R)$, and hence quite remote from the electroweak/confinement scales. 
Other possible locations of $v_{PQ}$ with respect to $H_I$ and $T_R$ are also possible.

In both solutions the Peccei-Quinn scalar field $\Phi$, displays an original symmetry which can be broken by gravitational effects, with a physical Goldstone mode $a(x)$ which remains such from the 
large $v_{PQ}$ scale down to $\Lambda_{QCD}$, when axion oscillations occur. 
In the DFSZ solution, the axion emerges as a linear combination of the phases of the CP-odd sector and of $\Phi$ which are orthogonal to the hypercharge $(Y)$ and are fixed by the normalization of the kinetic term of the axion field $a$. 
The solution to the strong CP problem is then achieved by rendering the parameter of the $\theta$-vacuum dynamical, with the angle $\theta$ replaced by the axion field ($\theta\to 
 a/f_a$), with $f_a$ being the axion decay constant. 
 
The computation of the axion mass $m_a$ is then derived from the vacuum energy of the $\theta$-vacuum 
$E(\theta)$ once this is re-expressed in terms of the QCD chiral Lagrangian, which in the 
two quark flavour (u,d) case describes the spontaneous breaking of the $SU(2)_L\times SU(2)_R$ flavour symmetry to a diagonal $SU(2)$ subgroup, with the 3 Goldstone modes $(\pi^\pm, \pi^0)$ 
being the dynamical field of the low energy dynamics. In this effective chiral description in which the $\theta$ parameter is present, 
the vacuum energy acquires a dependence both on neutral pseudoscalar $\pi^0$ and on $\theta$ of the form 
\beq
E(\pi^0,\theta)=- m_\pi^2 f_\pi^2  \sqrt{\cos^2\frac{\theta}{2} +\left(\frac{ m_d - m_u}{m_d + m_u}\right)^2 \sin^2\frac{\theta}{2}}\cos\left( \pi^0-\phi(\theta)\right)
\eeq
with 
\beq
\phi(\theta)\equiv \frac{m_d -m_u}{m_d +m_u}\sin\frac{\theta}{2}.
\eeq
At the minimum, when $\pi^0= f_\pi \phi(\theta)$, the vacuum energy assumes the simpler form 
\beq
E(\theta)=-m_\pi^2 f_\pi^2 \sqrt{1- \frac{4 m_u m_d}{(m_u + m_d)^2}\sin^2\frac{\theta}{2}}
\eeq
which expanded for small $\theta$ gives the well-know relation 
\beq
E(\theta)=-m_\pi^2 f_\pi^2 +\frac{1}{2} m_\pi^2 f_\pi^2 \frac{m_u m_d}{(m_u + m_d)^2}\theta^2 + \ldots
\eeq
and the corresponding axion mass 
\beq
m_a^2=\frac{m_\pi^2 f_\pi^2}{f_a^2} \frac{m_u m_d}{(m_u + m_d)^2}
\eeq
as $\theta\to a/f_a$. Before getting into a more detailed analysis of the various possible extensions of the traditional PQ scenarios, we briefly review the KSVZ (hadronic) and DFSZ (invisible) axion solutions. 
\subsection{KSVZ and DFSZ axions} 
In both the DFSZ and KSVZ scenarios a global anomalous $U(1)_{PQ}$ symmetry gets broken at some large scale $v_{PQ}$, with the generation of a Nambu-Goldstone mode from the 
CP-odd scalar sector. In the KSVZ case the theory includes a heavy quark $Q$ which acquires a large mass by a Yukawa coupling with the scalar $\Phi$. In this case the Lagrangian of $Q$ takes the form 

\beq
\mathcal{L}=\vline \partial \Phi\vline^2 +i \bar{Q}\slash{D}Q +  \lambda \Phi\bar{Q}_L Q_R + h.c.  - V(\Phi)
\eeq
with a global $U(1)_{\small PQ}$ chiral symmetry of the form 
\beqa
\Phi &\to & e^{i \alpha } \Phi \nn\\
Q &\to& e^{-\frac{i}{2} \alpha \gamma_5 }Q
\eeqa
with an $SU(3)_c$ covariant derivative ($D$) containing the QCD color charge of the heavy fermion $Q$.  The scalar PQ potential can be taken of the usual Mexican-hat form and it is $U(1)_{PQ}$ symmetric. Parameterising the PQ field with respect to its broken vacuum 
\beq
\Phi= \frac{\phi+ v_{\small {pq}}}{\sqrt{2}}e^{i \frac{a(x)}{v_{PQ}} }+\ldots
\eeq
the Yukawa coupling of the heavy quark $Q$ to the CP-odd phase of $\Phi$, $a(x)$, takes the form 
\beq
\lambda \frac{ v_{pq}}{\sqrt{2}}e^{i\frac{a(x)}{v_{PQ}}} \bar{Q}_L Q_R.
\eeq 
At this stage one assumes that there is a decoupling of the heavy quark from the low energy spectrum 
by assuming that $v_{PQ}$ is very large. The standard procedure in order to extract the low energy 
interaction of the axion field is to first redefine the field $Q$ on order to remove the exponential with the axion in the 
Yukawa coupling
\beq
 e^{i \gamma_5\frac{a(x)}{2 v_{PQ}}}Q_{L/R}\equiv Q'_{L/R}.
 \eeq
 This amounts to a chiral transformation which leaves the fermionic measure non-invariant 
\beq
\textrm{D}\bar{Q}\textrm{D}Q\to e^{i \int d^4 x \frac{6 a(x)}{32 \pi^2 v_{PQ}} G(x)\tilde G (x)}\textrm{D}\bar{Q}\textrm{D}Q
\eeq
and generates a direct coupling of the axion to the anomaly $G \tilde{G}$. Here the factor of 6 is related to the number of L/R components being rotated, which is 6 if $Q$ is assigned to the triplet of $SU(3)_c$.

The kinetic term of $Q$ is not invariant under this field redefinition and generates a derivative coupling of $a(x)$ to the axial vector current of $Q$. For $n_f$ triplets, for instance, the effective action of the axion, up to dimension-5 takes the form 
\beq
\mathcal{L}_{eff}=\frac{1}{2}\partial_\mu a(x)\partial^\mu a(x) + \frac{6 n_f}{32 \pi^2 v_{PQ}}a(x) G \tilde G + \frac{1}{v_{PQ}}\partial_\mu a \bar{Q} \gamma^\mu \gamma_5 Q +\ldots
\eeq
where we have neglected extra higher dimensional contributions, suppressed  by $v_{PQ}$.\\
In the case of the DFSZ axion, the solution to the strong CP problem is found by introducing a scalar $\Phi$ together with two Higgs doublets $H_u$ and $H_d$. 
In this case one writes down a general potential, function of these three fields, which is $SU(2)\times U(1)$ invariant and possesses a global symmetry 
\beq
H_u\to e^{i\alpha X_u }H_u,\qquad  H_d\to e^{i\alpha X_d }H_d, \qquad \Phi\to e^{i\alpha X_\Phi}\Phi
\eeq
with  $X_u +X_d=-2 X_\Phi$. It is given by a combination of terms of the form
\beq
\label{mixing}
V= V( \vline H_u\vline^2 \,,\, \vline H_d\vline^2\, ,\,\vline \Phi\vline^2 \,,\,\vline H_u H_d^\dagger \vline^2\, , \,\vline H_u\cdot H_d\vline^2 \, ,\, H_u\cdot H_d, \Phi^2 ) 
\eeq
where $H_u\cdot H_d$ denotes the $SU(2)$ invariant scalar product.
The identification of the axion field is made by looking for a linear combination of the phases which is not absorbed by a gauge transformation. This can be done, for instance, 
by going to the unitary gauge and removing all the NG modes of the broken gauge symmetry. The corresponding phase, which is the candidate axion, 
is the result of a process of mixing of the PQ field with the Higgs sector at a scale where the symmetry of the potential is spontaneously broken by the two Higgs fields. 

\section{ TeV scale: St\"uckelberg axions in anomalous $U(1)$ extensions of the Standard Model} 

Intersecting D-brane models are one of those constructions where generalized axions appear \cite{Kiritsis:2003mc,Ibanez:2001nd, Antoniadis:2002qm,Blumenhagen:2006ci}. 
In the case in which several stacks of such branes are introduced, each stack being the domain in which fields with the gauge symmetry $U(N)$ live, several intersecting stacks generate
at their common intersections, fields with the quantum numbers of all the unitary gauge groups of the construction, such as 
\beq
U(N_1)\times U(N_2)\times ...\times U(N_k)=SU(N_1)\times U(1)\times SU(N_2)\times U(1)\times  ...\times SU(N_k)\times U(1). 
\eeq
The phases of the extra $U(1)$'s are rearranged in terms of an anomaly-free generator, corresponding to an (anomaly free) hypercharge $U(1)$ (or $U(1)_Y$), times extra 
$U(1)$'s which are anomalous, carrying both their own anomalies and the mixed anomalies with all the gauge factors of the Standard Model.  This general construction can be made phenomenologically interesting.

Using this approach, the Standard Model can be obtained by taking for example 3 stacks of branes: a first stack of 3 branes, yielding a $U(3)$ gauge symmetry, a second stack of 2 branes, yielding a symmetry 
$U(2)$ and an extra single $U(1)$ brane, giving a gauge structure of the form $SU(3)\times SU(2)\times 
U(1)\times U(1)\times U(1)$. Linear combinations of the generators of the three $U(1)$'s allow us to rewrite the entire abelian symmetry in the form $U(1)_Y\times U(1)' \times U(1)''$, with the remaining $U(1)' \times U(1)''$ factors carrying anomalies which need to be cancelled 
by extra operators. 
The simplest realization of the Standard Models (SM) is obtained by 2 stacks and a single brane at their intersections, giving a symmetry $U(3)\times U(2)\times U(1)$. In this case, in the hypercharge basis, 
the gauge structure of the model can be rewritten in the form $SU(3)_c\times SU(2)_w\times U(1)_Y\times U(1)' \times U(1)''$. 

We consider the case of a single $U(1)'\equiv U(1)_B$ anomalous gauge symmetry, where the St\"uckelberg field $b(x)$ couples to the gauge field $B_\mu$ by the gauge invariant term
\ba
\mathcal{L}_{St}=\frac{1}{2}\left( \partial_\mu b - M B_\mu\right)^2 
\label{stuclag}
\ea
which is the well-known St\"uckelberg form. $M$ is the St\"uckelberg mass. The St\"uckelberg symmetry of the Lagrangian (\ref{stuclag}) is revealed by acting with gauge transformations of the gauge fields $B_\mu$ under which the axion $b$ varies by a local shift 
\ba
&&\delta_B B_\mu=\partial_\mu \theta_B \qquad \qquad \delta b= M \theta_B  
\ea
parameterized by the local gauge parameters $\theta_B$. Originally, the St\"uckelberg symmetry was presented as a way to give a mass to an abelian gauge field while still preserving the gauge invariance of the theory. However, it is clear nowadays that its realization is 
the same one as obtained, for instance, in an abelian-Higgs model when one decouples the radial excitations of the Higgs fields from its phase \cite{Coriano:2009zh}. The bilinear $\partial B b $ mixing present in Eq. \eqref{stuclag} is an indication that the $b$ field describes a Nambu-Goldstone mode which could, in principle, be removed by a unitary gauge condition. We will come back to this point later in this review.
There is a natural way to motivate Eq. \eqref{stuclag}.

If we assume that the $U(1)_B$ gauge symmetry is generated within string theory and realized around the Planck scale, the massive anomalous gauge boson acquires a mass through the presence of an $A\wedge F\,$ coupling in the bosonic sector of a string-inspired effective action \cite{Ghilencea:2002da}. The starting Lagrangian of the effective theory involves, in this case, an antisymmetric  rank-2 tensor $A_{\mu\nu}$ coupled to the field strength $F_{\mu\nu}$ of  $B_\mu$
\begin{equation}
\label{dd} 
{\cal L}\ =\ -\frac{1}{12} H^{\mu\nu\rho}
H_{\mu\nu\rho}-\frac{1}{4g^2} F^{\mu\nu} F_{\mu\nu} 
+ \frac{M}{4}\ \epsilon^{\mu\nu\rho\sigma} { A}_{\mu\nu}\ F_{\rho\sigma},
\end{equation}
where 
\beq
\label{field3}
H_{\mu\nu\rho}=\partial_\mu A_{\nu\rho}+\partial_\rho A_{\mu\nu}
+\partial_\nu A_{\rho\mu}, \qquad  F_{\mu\nu}=\partial_\mu
B_\nu-\partial_\nu B_\mu
\eeq
is the kinetic term for the 2-form and $g$
is an arbitrary constant. Besides the two kinetic terms for $A_{\mu\nu}$ and $B_\mu$, the third contribution in Eq.~(\ref{dd}) is the $A\wedge F$ interaction. 

The Lagrangian is dualized by using a ``first order'' formalism, where $H$ is treated independently from the antisymmetric field $A_{\mu\nu}$. This is obtained by introducing a constraint with a 
Lagrangian multiplier field $b(x)$ in order to enforce the condition $H=dA$ from the equations of motion of $b$, in the form  
\begin{equation}
\label{secondform}
{\cal L}_0=-\frac{1}{12} H^{\mu\nu\rho} H_{\mu\nu\rho}-\frac{1}{4g^2} F^{\mu\nu}\ F_{\mu\nu} 
- \frac{M}{6}\ \epsilon^{\mu\nu\rho\sigma} H_{\mu\nu\rho}\ B_{\sigma} 
+\frac{1}{6}\,b(x)\,\epsilon^{\mu\nu\rho\sigma} \partial_\mu H_{\nu\rho\sigma}.
\end{equation}
The appearance of a scale $M$ in this Lagrangian is crucial for the cosmological implications of such a theory \cite{Armillis:2008vp}, since it defines the energy region where the mechanism of anomaly 
cancellation comes into play \cite{Coriano:2005js}. The last term in (\ref{secondform}) is necessary in order to reobtain (\ref{dd}) from (\ref{secondform}). If, instead, we integrate by parts the last term of the Lagrangian given in (\ref{secondform}) and solve trivially for $H$ we find 
\begin{equation}
H^{\mu\nu\rho}= - \epsilon^{\mu\nu\rho\sigma}\left(M B_\sigma-\partial_\sigma b\right),
\end{equation}
and inserting  this result back into (\ref{secondform}) we obtain the expression
\begin{equation}
\label{eee}
{\cal L}_{A}\ =\ -\frac{1}{4g^2}\ F^{\mu\nu}\ F_{\mu\nu} - \frac{1}{2} \left(M B_\sigma-\partial_\sigma b\right)^2
\end{equation}
which is the St\"uckelberg form for the mass term of $B$. This rearrangement of the degrees of freedom is an example of the connection between Lagrangians of antisymmetric tensor fields 
and their dual formulations which, in this specific case, is an abelian massive Yang-Mills theory in a St\"uckelberg form.\\
The axion field, generated by the dualization mechanism, appears as a Nambu-Goldstone mode, which can be removed by a unitary gauge choice. However, as discussed in \cite{Coriano:2005js}, 
the appearance, at a certain scale, of an extra potential which will mix this mode with the scalar sector, will allow to extract a physical component out of $b$, denoted by $\chi$.

 The origin of such a mixing potential is here assumed to be of non-perturbative origin and triggered at a scale below the St\"uckelberg scale $M$. It is at this second scale where a physical axion appears 
 in the spectrum of the theory. The local shift invariance of $b(x)$ is broken by the vev of the Higgs sector appearing in the part of the potential that couples the St\"uckelberg field to the remaining scalars, causing a component of the St\"uckelberg to become physical. 
 The scale at which this second potential is generated and gets broken is the second scale controlling the mass of the axion, $\chi$. Such a potential is by construction periodic in $\chi$, as we are going to illustrate below and it is quite similar to the one discussed in Eq. (\ref{mixing}).
Its size is controlled by constants ($\lambda_i$) which are strongly suppressed by the exponential factor  ($\sim e^{-S_{inst}}$, with $S_{inst}$ the instanton action), determined by the value of the action in the instanton background. 

In models with several $U(1)$'s  this construction is slightly more involved, but the result of the mixing of the CP odd phases leaves as a remnant, also in this case, only one physical axion \cite{Coriano:2005js}, whose mass is controlled by the size of the Higgs-axion mixing. 
\begin{table}[t]
\begin{center}
\begin{tabular}{|c|c|c|c|c|c|c|}
\hline
$ f $ & $Q$ &  $ u_R $ &  $ d_R $ & $ L $ & $e_R$ \\
\hline \hline
$q^B$  &  $q^B_Q$  & $q^B_{u_R}$  &  $q^B_{d_R}$ & $q^B_L$ & $q^B_{e_R}$ \\ \hline
\end{tabular}
\end{center}
\centering
\renewcommand{\arraystretch}{1.2}
\begin{tabular}{|c|c|c|c|c|}\hline
$f$ & $SU(3)^{}_C$ & $SU(2)^{}_L$ & $U(1)^{}_Y$ & $U(1)^{}_B$ \\ \hline \hline
$Q$ &  3 & 2 & $ 1/6$ & $q_Q^B$\\
$u_R$ &  3 & 1 & $ 2/3$ & $q_Q^B+q^B_{u}$\\
$d_R$ &  3 & 1 & $ -1/3$ & $q_Q^B-q^B_{d}$\\
$L$ &  1 & 2 & $ -1/2$ & $q^B_L $\\ 
$e_R$ &  1 & 1 & $-1$ & $ q^B_L - q^B_{d}$\\ 
\hline
$H^{}_u$ &  1 & 2 & $1/2$ & $ q^B_u $\\ 
$H^{}_d$ &  1 & 2 & $1/2$ & $ q^B_d  $\\ 
\hline
\end{tabular}
\caption{\small Charges of the fermion and of the scalar fields \label{solve_q}}
\label{Table}
\end{table}

\subsection{St\"uckelberg models at the TeV scale with two-Higgs doublets }

The type of models investigated in the past have been  formulated around the TeV scale and discussed in detail in their various sectors  
\cite{Coriano:2006xh,Coriano:2007fw,Coriano:2007xg,Armillis:2008vp,Coriano:2008pg,Coriano:2009zh,Armillis:2008bg} \cite{Frampton:1996cc}. We offer a brief description of such realizations, 
which extend the symmetry of the SM minimally and as such are simpler than in other realizations involving larger gauge symmetries. 
 They have the structure of effective actions where dimension-5 interactions are introduced in order to restore the gauge invariance of the Lagrangian in the presence of an 
 anomalous gauge boson (and corresponding fermion spectrum). Therefore, they are quite different from ordinary anomaly-free versions of the same theories. They include one extra anomalous $U(1)_B$ symmetry, the St\"uckelberg field and a set of 
scalars with a sufficiently wide CP odd sector in order to induce a mixing potential between the scalar fields and the St\"uckelberg. Obviously, such models are of interest at the 
LHC for predicting anomalous gauge interactions in the form of extra neutral currents \cite{Armillis:2008vp,Armillis:2007tb} with respect to those of the electroweak sector.  

\begin{figure}[t]
\begin{align}
S_{eff}=&S_0+
\begin{minipage}[c]{70pt}
\includegraphics[scale=.7]{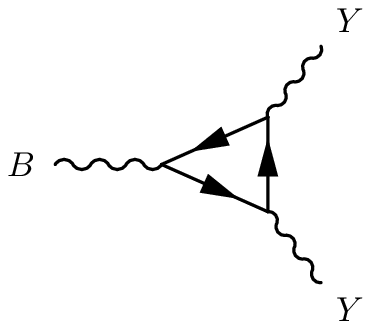}
\end{minipage}+
\begin{minipage}[c]{70pt}
\includegraphics[scale=.7]{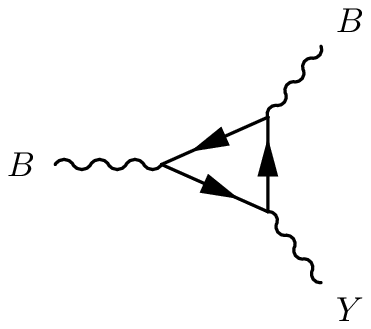}
\end{minipage}+
\begin{minipage}[c]{70pt}
\includegraphics[scale=.7]{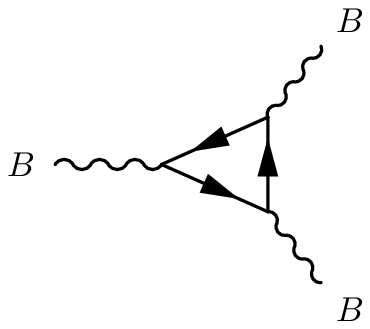}
\end{minipage}+
\begin{minipage}[c]{70pt}
\includegraphics[scale=.7]{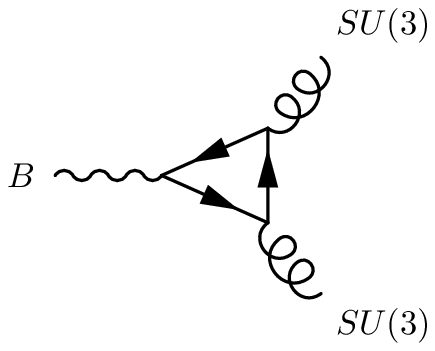}
\end{minipage}+
\begin{minipage}[c]{70pt}
\includegraphics[scale=.7]{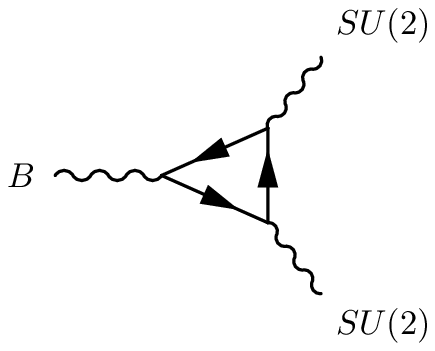}
\end{minipage}+\nonumber\\
&\nonumber\\
&
\begin{minipage}[c]{70pt}
\includegraphics[scale=.7]{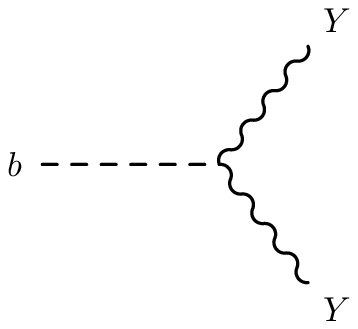}
\end{minipage}+
\begin{minipage}[c]{70pt}
\includegraphics[scale=.7]{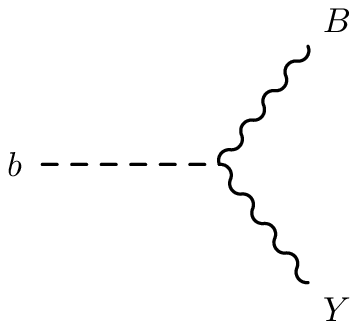}
\end{minipage}+
\begin{minipage}[c]{70pt}
\includegraphics[scale=.7]{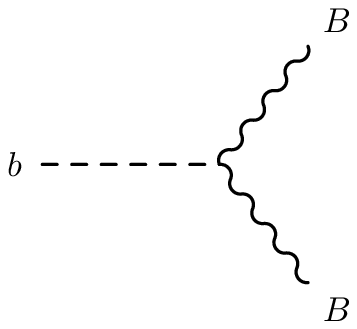}
\end{minipage}+
\begin{minipage}[c]{70pt}
\includegraphics[scale=.7]{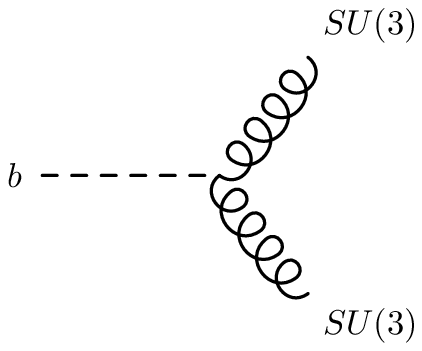}
\end{minipage}+
\begin{minipage}[c]{70pt}
\includegraphics[scale=.7]{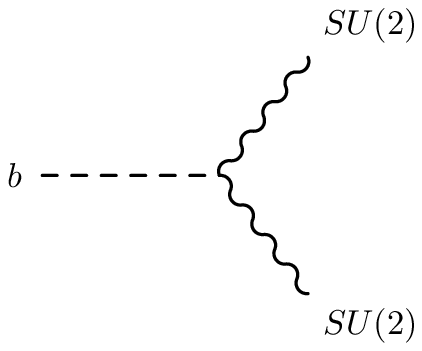}
\end{minipage}\nonumber
\end{align}
\caption{The 1PI effective action for a typical low scale model obtained by adding one extra anomalous $U(1)_B$ to the Standard Model action. Shown are the one-loop trilinear anomalous interactions and the corresponding counterterms, involving the $b$ field.} 
\label{fig:lagrangian}
\label{eaction}
\end{figure}

The effective action has the structure given by
\beqn
{\mathcal S} &=&   {\mathcal S}_0 + {\mathcal S}_{Yuk} +{\mathcal S}_{an} + {\mathcal S}_{WZ} 
\label{defining}
\eeqn
where ${\mathcal S}_0$ is the classical action. The same structure will characterize also other, more complex, realizations. 
It contains the usual gauge degrees of freedom of the Standard Model plus the extra anomalous gauge boson $B$ which is already massive before electroweak symmetry breaking, via a St\"uckelberg mass term, as it is clear from \eqref{eee}.
We show the structure of the 1-particle irreducible effective action in Fig. \ref{eaction}. We consider a 2-Higgs doublet model for definiteness, which will set the ground for more complex extensions that we will address in the next sections. We consider 
an $SU(3)_c\times SU(2)_w\times U(1)_Y\times U(1)_B$ gauge symmetry model, characterized by an action $\mathcal{S}_0$, corresponding to the first contribution shown in Fig.~\ref{eaction}, 
plus one loop corrections which are anomalous and break gauge invariance whenever there is an insertion of the anomalous gauge boson $B_\mu$ in the trilinear fermion vertices. In the last line of the same figure are shown the 
$(b/M) F\wedge F$ Wess-Zumino counterterms needed for restoring gauge invariance, which are suppressed by the St\"uckelberg scale $M$.
 Table \ref{Table} shows the charge assignments of the fermion spectrum of the model, where we have indicated by $q$ the charges for a single generation, having taken into account 
 the conditions of gauge invariance of the Yukawa couplings. Notice that the two Higgs fields carry different charges under $U(1)_B$, which allow to extend the ordinary scalar potential of the 
 two-Higgs doublet model by a certain extra contribution. This will be periodic in the axi-Higgs $\chi$, after the two Higgses, here denoted as $H_u$ and $H_d$, acquire a vev. 
Specifically, $ q^B_L, q^B_Q$ denote the charges of the left-handed lepton doublet $(L)$ and of the quark doublet $(Q)$ respectively, while $q^B_{u_r},q^B_{d_r}, q^B_{e_R}$ are the charges of the 
right-handed $SU(2)$ singlets (quarks and leptons). We denote by $\Delta q^B=q^B_u - q^B_d$ the difference between the two charges of the up and down Higgses $(q^B_u, q^B_d)$ respectively and from now on we will assume that it is non-zero.
The trilinear anomalous gauge interactions induced by the anomalous $U(1)$ and the relative counterterms, which are all parts of the 1-loop effective action, are illustrated in Fig. \ref{fig:lagrangian}. 

\subsection{Fermion/gauge field couplings}

The models that we are discussing are characterized by one extra neutral current, mediated by a $Z'$ gauge boson. The interaction of the fermions with the gauge fields is defined by the Lagrangian
\begin{align}
{\cal L}_{int}^{quarks} =&   
\begin{pmatrix} 
\bar{ u }_{L\,i} & \bar{ d }_{L\,i} 
\end{pmatrix}
\gamma^{\mu}
\left[-g_s T^a G^a_\mu- g_2 \tau^a W^a_\mu-\frac{1}{12}g_Y  Y_\mu-\frac{1}{2}g_B q^B_Q B_\mu \right]  
\begin{pmatrix} 
u_{L\,i} \cr 
d_{L\,i}
\end{pmatrix}  + \nonumber\\
& + \bar{u}_{R\,i} \gamma^{\mu} 
\left[-g_s T^a G^a_\mu- g_2 \tau^a W^a_\mu-\frac{1}{3}g_Y Y_\mu-\frac{1}{2}g_B q^B_{u_R} B_\mu \right]  u_{R\,i}  \nonumber\\
& + \bar{d}_{R\,i} \; {\gamma}^{\mu}
\left[-g_s T^a G^a_\mu- g_2 \tau^a W^a_\mu+\frac{1}{6}g_Y Y_\mu-\frac{1}{2}g_B q^B_{d_R} B_\mu\right] d_{R\,i}.
\end{align}
while the Higgs sector  is characterized by the two Higgs doublets 
\ba
H_u=\left(
\begin{array}{c}
H_u^+\\
H_u^0
\end{array}\right) \qquad 
H_d = \left(
\begin{array}{c}
H_d^+\\
H_d^0 
\end{array}\right)
\ea
where $H_u^+$, $H_d^+$ and $H_u^0$, $H_d^0$ are complex fields with (with some abuse of notation we rescale the fields by a factor of $1/\sqrt{2}$)
\ba
H_u^+ =  \frac{\textrm{Re}H_{u}^+ + i \textrm{Im}H_{u}^+}{\sqrt{2}} ,\qquad
H_d^- =  \frac{\textrm{Re}H_{d}^- + i \textrm{Im}H_{d}^-}{\sqrt{2}} , \qquad
H_u^- = H_u^{+ *}, \qquad
H_d^+ = H_d^{- *}.
\ea
Expanding around the vacuum we get for the neutral components
\ba
H_u^0 =  v_u + \frac{\textrm{Re}H_{u}^0 + i \textrm{Im}H_{u}^0}{\sqrt{2}} , \qquad
H_d^0 =  v_d + \frac{\textrm{Re}H_{d}^0 + i \textrm{Im}H_{d}^0}{\sqrt{2}}. 
\label{Higgsneut}
\ea
which will play a key role in determining the mixing of the St\"uckelberg field in the periodic potential.
The electroweak mixing angle is defined by $\cos\theta_W= g_2/g, \sin\theta_W= g_Y/g$, with $g^2= g_Y^2 + g_2^2$.
We also define $\cos \beta=v_d/v$, $ \sin \beta=v_u/v$ with $v^2=v_d^2 + v_u^2$.
The matrix rotates the neutral gauge bosons from the interaction to the mass eigenstates after electroweak symmetry breaking and has elements which are $O(1)$, being expressed in terms of ratios of coupling constants, which correspond to mixing angles. It is given by
\ba
\begin{pmatrix}A_\g \cr Z \cr Z^{\prime} 
\end{pmatrix}\,=O^A \begin{pmatrix} W_3 \cr A^Y \cr B
\end{pmatrix}
\label{OA}
\ea
which can be approximated to leading order as
\bea
O^A  \simeq  \begin{pmatrix}
\frac{g^{}_Y}{g}           &     \frac{g^{}_2}{g}         &      0   \cr
\frac{g^{}_2}{g} + O(\epsilon_1^2)          &     -\frac{g^{}_Y}{g} + O(\epsilon_1^2) &      \frac{g}{2} \epsilon_1 \cr
-\frac{g^{}_2}{2}\epsilon_1     &     \frac{g^{}_Y}{2}\epsilon_1  &   1 + O(\epsilon_1^2)
\end{pmatrix}
\label{matrixO}
\eea
where
\begin{align}
&\epsilon_1=\frac{x_B}{M^2},\nonumber\\
&x_B=\left(q^{B}_u v_u^2 + q^{B}_d v_d^2\right).
\end{align}
Once the WZ counterterms will be rotated into the gauge eigenstates and the $b$ field into the physical $\chi$ field, there will be a direct coupling of the anomaly to the physical gauge bosons. 
This will involve both the neutral and the charged sectors. More details can be found in \cite{Coriano:2007xg}. 

\subsection{Counterterms}

Fixing the values of the counterterms in simple single $U(1)$ models like the one we are reviewing, allows to gain some insight into 
the possible solutions of the gauge invariance conditions on the Lagrangian. 
The numerical values of the counterterms appearing in the second line of Fig.~\ref{fig:lagrangian} are fixed by such conditions, giving
\begin{align}
&C_{BYY} =  -\frac{1}{6}q^B_Q+\frac{4}{3}q^B_{u_R}+\frac{1}{3}q^B_{d_R}-\frac{1}{2}q^B_L+q^B_{e_R},
\nonumber\\
&C_{YBB} = -(q^{B}_Q)^2+2 (q^{B}_{u_r})^2-(q^{B}_{d_R})^2+(q^{B}_L)^2-(q^{B}_{e_R})^2,
\nonumber\\
&C_{BBB} = -6 (q^B_Q)^3+3 (q^B_{u_R})^3+3 (q^B_{d_R})^3-2 (q^B_L)^3+(q^B_{e_R})^3,\nonumber
\end{align}
\begin{align}
\nonumber\\
&C_{Bgg}=\frac{1}{2} (-2 q^B_Q+q^B_{d_R}+q^B_{u_R}),
\nonumber\\
&C_{BWW} = \frac{1}{2} (-q^B_L-3 q^B_Q).
\end{align}
They are, respectively, the counterterms for the cancellation of the mixed anomaly $U(1)_B U(1)_Y^2$ and $U(1)_Y U(1)_B^2$; the 
counterterm for the $BBB$ anomaly vertex or $U(1)_B^3$ anomaly, and those of the $U(1)_BSU(3)^2$ and $U(1)_B SU(2)^2$ anomalies. They are defined in the Appendix. From the Yukawa couplings we get the following constraints on the $U(1)_B$ charges
\ba
q^B_Q-q^B_{d}-q^B_{d_R}=0\hspace{1cm}
q^B_Q+q^B_{u}-q^B_{u_R}=0\hspace{1cm}
q^B_L-q^B_{d}-q^B_{e_R}=0.
\ea
Using the equations above, we can eliminate some of the charges in the expression of the counterterms, obtaining
\begin{align}
&C_{BYY} = \frac{1}{6} (3 q^B_L+9 q^B_Q+8 \Delta q^B),
\nonumber\\
&C_{YBB} = 2 \left[q^B_d (q^B_L+3 q^B_Q)+2 \Delta q^B (q^B_d+q^B_Q)+(\Delta q^B)^2\right],
\nonumber\\
&C_{BBB} = (q^B_L-q^B_d)^3+3 (q^B_d+q^B_Q+\Delta q^B)^3+3 (q^B_Q-q^B_d)^3-2 (q^B_L)^3-6 (q^B_Q)^3,
\nonumber\\
&C_{Bgg}=\frac{\Delta q^B}{2},
\nonumber\\
&C_{BWW} = \frac{1}{2} (-q^B_L-3 q^B_Q).
\label{charge_asynew}
\end{align}
The equations above parametrize, in principle, an infinite class of models whose charge assignments under $U(1)_B$ are arbitrary, with the charges in the last column of Tab. (\ref{solve_q}) taken as their free parameters. 
The coupling of the axion to the corresponding gauge bosons can be fixed by a complete solution to the anomaly constraints, which may provide us with an insight into the possible mechanisms of misalignment that could take place at both the electroweak and at the QCD phase transitions.

\subsection{Choice of the charges}

Due to the presence, in general, of a nonvanishing mixed anomaly of the $U(1)_B$ gauge factor with both $SU(2)$ and $SU(3)$, the St\"uckelberg axion of the model has interactions with both the strong and the weak sectors, which both support instanton solutions, and therefore could acquire a mass non-perturbatively both at the electroweak and at the QCD phase transitions. In this case we take into account the possibility of having sequential misalignments, with the largest contribution to the mass coming from the latter.  
Obviously, for a choice of charges characterized by $\Delta q=0$, in which both doublets of the Higgs 
sector $H_u$ and $H_d$ carry the same charge under $U(1)_B$, the axion mass will not acquire any instanton correction at the QCD phase transition. In this case the potential responsible for Higgs-axion mixing would vanish.
In this scenario a solution to the anomaly equations with a vanishing electroweak interaction of the St\"uckelberg can be obtained by choosing $q^B_L= -3 q^B_Q$. 

If instead the charges are chosen in a way to have both non-vanishing weak ($C_{BWW}$) and strong ($C_{Bgg}$) counterterms, it is reasonable to expect that the misalignment of the axion potential will be sequential, 
with a tiny mass generated at the electroweak phase transition, followed by a second misalignment induced at the strong phase transition. The instanton configurations of the weak and strong sectors will be contributing differently to the mass of the physical axion.
However, due to the presence of a coupling of this field with the strong sector, its mass will be significantly dominated by the QCD phase transition, as in the Peccei-Quinn case.\subsection{The scalar sector}
The scalar sector of the anomalous abelian models  is characterized, as already mentioned, by the ordinary electroweak potential of the SM involving, in the simplest formulation, 
two Higgs doublets $V_{P Q}(H_u, H_d)$ plus one extra contribution, denoted as $V_{\slashed{P}\slashed{Q}}(H_u,H_d,b)$ - or $V^\prime$ (PQ breaking) in 
\cite{Coriano:2005js} - which mixes the Higgs sector with the St\"uckelberg axion $b$, needed for the restoration of the gauge invariance of the effective Lagrangian
\begin{equation}
\label{pot}
V=V_{PQ}(H_u,H_d) + V_{\slashed{P}\slashed{Q}}(H_u,H_d,b).
\end{equation}
The appearance of the physical axion in the spectrum of the model takes place after 
the phase-dependent terms - here assumed to be of non-perturbative origin and generated at a phase transition - 
find their way in the dynamics of the model and induce a curvature on the scalar potential. The mixing induced in the CP-odd sector determines the presence 
of a linear combination of the St\"uckelberg field $b$ and of the Goldstones of the CP-odd sector which acquires a tiny mass. From (\ref{pot}) we have a first term
\ba
V_{PQ}&=&\mu_u^2 H_u^{\dagger}H_u+\mu_d^2 H_d^{\dagger}H_d+\lambda_{uu}(H_u^{\dagger}H_u)^2
+\lambda_{dd}(H_d^{\dagger}H_d)^2-2 \lambda_{ud}(H_u^{\dagger}H_u)(H_d^{\dagger}H_d)
+2\lambda^{\prime}_{ud}\vert H_u^T \tau_2 H_d\vert^2 \nonumber \\
\eeqa
typical of a two-Higgs doublet model, to which we add a second PQ breaking term
\ba
V_{\slashed{P}\slashed{Q}}&=&\lambda_0(H_u^{\dagger}H_d e^{-i g_B (q_u-q_d)\frac{b}{2 M}})+
\lambda_1(H_u^{\dagger}H_d e^{-i g_B (q_u-q_d)\frac{b}{2 M}})^2+\lambda_2(H_u^{\dagger}H_u)(H_u^{\dagger}H_d e^{-i g_B (q_u-q_d)\frac{b}{2M}})+\nonumber\\
&&\lambda_3(H_d^{\dagger}H_d)(H_u^{\dagger}H_d e^{-i g_B (q_u-q_d)\frac{b}{2 M}})+\textrm{h.c.}
\ea
These terms are allowed by the symmetry of the 
model and are parameterized by one dimensionful  ($\lambda_0$) and three dimensionless couplings ($\lambda_1,\lambda_2,\lambda_3$).  
Their values are weighted by an exponential factor containing as a suppression the instanton action. In the equations below we will rescale 
$\lambda_0$ by the electroweak scale $v=\sqrt{v_u^2 + v_d^2}$  ($\lambda_0 \equiv \bar{\lambda}_0 v$) so as to obtain a homogeneous expression for the mass of $\chi$ 
as a function of the relevant scales of the model which are, besides the electroweak vev $v$ the St\"uckelberg mass $M$ and the anomalous gauge coupling of the $U(1)_B$, $g_B$.

The gauging of an anomalous symmetry has some important effects on the properties of this pseudoscalar, first among all the appearance of independent mass and couplings 
to the gauge fields. This scenario allows then a wider region of parameter space in which one could look for such particles \cite{Coriano:2006xh,Coriano:2007xg,Coriano:2009zh},
rendering them "axion-like particles" rather than usual axions. We will still refer to them as axions for simplicity.
So far only two complete models have been put forward for a consistent analysis of these types of particles, the first one non-supersymmetric  \cite{Coriano:2005js} and a second one supersymmetric \cite{Coriano:2008xa}.

\subsection{The potential for a generic St\"uckelberg mass} 
 
The physical axion $\chi$ emerges as a linear combination of the phases of the various complex scalars appearing in combination with the $b$ field. 
To illustrate the appearance of a physical direction in the phase of the extra potential, we focus our attention on just the CP-odd sector of the total potential, 
which is the only one that is relevant for our discussion.  The expansion of this potential around the electroweak vacuum is given by the parameterization 
\ba
H_u=\left(
\begin{tabular}{c}
$H_u^+$\\
$v_u+H_u^0$
\end{tabular}
\right)
\hspace{1cm}
H_d=\left(
\begin{tabular}{c}
$H_d^+$\\
$v_d+H_d^0$
\end{tabular}
\right).
\ea
where $v_u$ and $v_d$ are the two vevs of the Higgs fields.
This potential is characterized by two null eigenvalues corresponding to two neutral Nambu-Goldstone modes $(G_0^1,G_0^2)$
and an eigenvalue corresponding to a massive state with an axion component ($\chi$). In the 
$(\textrm{Im}H_d^0,\textrm{Im} H_u^0, b)$ CP-odd basis we obtain the following normalized eigenstates
\begin{align}
G_0^1&=\frac{1}{\sqrt{v_u^2+v_d^2}}(v_d,v_u,0)\nonumber\\
G_0^2&=\frac{1}{\sqrt{g_B^2 (q_d-q_u)^2 v_d^2 v_u^2+2 M^2 \left(v_d^2+v_u^2\right)}}\left(-\frac{ g_B (q_d-q_u)v_d v_u^2}{\sqrt{v_u^2+v_d^2}},\frac{g_B (q_d-q_u)v_d^2 v_u}{\sqrt{v_d^2+v_u^2}},\sqrt{2}M\sqrt{v_u^2+v_d^2}\right)\nonumber\\
\chi&=\frac{1}{\sqrt{g_B^2 (q_d-q_u)^2 v_u^2 v_d^2+2M^2(v_d^2 + v_u^2)}}
\left(\sqrt{2} M v_u,-\sqrt{2} M v_d, g_B (q_d-q_u) v_d v_u\right)
\end{align}
and we indicate with $O^{\chi}$ the orthogonal matrix which allows to rotate them to the physical basis
\ba
\begin{pmatrix}
G_0^1 \cr
G_0^2 \cr
\chi
\end{pmatrix}
= O^\chi
\begin{pmatrix}
\textrm{Im}H^0_d \cr
\textrm{Im}H^0_u \cr
b
\end{pmatrix},
\ea
which is given by 
\begin{equation}
O^\chi=
\begin{pmatrix}
\frac{v_d}{v} & \frac{v_u}{v} & 0 \cr
-\frac{g_B (q_d-q_u)v_d v_u^2}{v\sqrt{g_B^2 (q_d-q_u)^2 v_d^2 v_u^2+2 M^2 v^2}}&
\frac{g_B (q_d-q_u)v_d^2 v_u}{v\sqrt{g_B^2 (q_d-q_u)^2 v_d^2 v_u^2+2 M^2 v^2}} &
\frac{\sqrt{2}M v}{\sqrt{g_B^2 (q_d-q_u)^2 v_d^2 v_u^2+2 M^2 v^2}} \cr
\frac{\sqrt{2} M v_u}{\sqrt{g_B^2 (q_d-q_u)^2 v_u^2 v_d^2+2M^2 v^2}}&
-\frac{\sqrt{2} M v_d}{\sqrt{g_B^2 (q_d-q_u)^2 v_u^2 v_d^2+2M^2 v^2}} &
\frac{g_B (q_d-q_u) v_d v_u}{\sqrt{g_B^2 (q_d-q_u)^2 v_u^2 v_d^2+2M^2 v^2}} 
\end{pmatrix}
\label{stella}
\end{equation}
where $v=\sqrt{v_u^2+v_d^2}$.\\
$\chi$ inherits WZ interaction since $b$ can be related to the physical axion $\chi$ and to the Nambu-Goldstone modes via this matrix as
\ba
b &=&  O_{13}^{\chi} G_0^1 + O_{23}^{\chi} G_0^2 + O_{33}^{\chi} \chi ,       
\label{rot12}
\ea
or, conversely,
\ba
\chi &=& O_{31}^{\chi} \textrm{Im}H_d + O_{32}^{\chi} \textrm{Im}H_u + O_{33}^{\chi} b.        
\ea
Notice that the rotation of $b$ into the physical axion $\chi$ involves a factor $O_{33}^{\chi}$ which is of order $v/M$.
This implies that $\chi$ inherits from $b$ an interaction with the gauge fields which is suppressed by a scale $M^2/v$. This scale is the product of two contributions: a $1/M$ suppression coming from the original Wess-Zumino counterterm of the Lagrangian ($b/M F\tilde{F}$) and a factor $v/M$ obtained by the projection of $b$ into $\chi$ due to $O_\chi$.\\
The direct coupling of the axion to the physical gauge bosons via the Wess-Zumino counterterms is obtained by the usual rotation to the mass eigenstates which can be obtained from the rotation matrix $O^A$ defined in \eqref{matrixO}. The final expression of the coupling of the axi-Higgs to the photon $g_{\chi\gamma\gamma}\chi F_\gamma\tilde{F_\gamma}$, 
is defined by a combination of matrix elements of the rotation matrices $O^A$ and $O^\chi$. Defining $g^2=g_2^2 + g_Y^2$, the expression of this coefficient can be derived in the form 
\beq
g^{\chi}_{\g \g}\,= \frac{g_B g_Y^2 g_2^2}{32 \pi^2 M g^2} O^{\chi}_{3\,3}\sum_f\left(- q^B_{f\,L} +q^B_{f \,R} \left(q^Y_{f \,R}\right)^2 - q^B_{f \,L} \left(q^Y_{f \,L}\right)^2\right).
 \label{gchi}
 \eeq 
Notice that this expression is cubic in the gauge coupling constants, since factors such as $g_2/g$ and $g_Y/g$ are mixing angles while the factor $1/\pi^2$ originates from the anomaly. Therefore  one obtains a general behaviour for $g^{\chi}_{\g \g}$ of $O(g^3 v/M^2)$, with charges which are, in general, of order unity. 
\subsection{Periodicity of the extra potential}
Equivalently, it is possible to reobtain the results above by an analysis of the phases of the extra potential, which shows how this becomes periodic in $\chi$, the axi-Higgs. This approach shows also quite directly the gauge invariance of $\chi$ as a physical pseudoscalar. In fact, if we opt for a polar parametrization of the neutral components in the broken phase 
\ba
H_u^0=\frac{1}{\sqrt{2}}\left(\sqrt{2}v_u + \rho_u^0(x) \right) e^{i\frac{F_u^0(x)}{\sqrt{2}v_u}}
\hspace{1cm}
H_d^0=\frac{1}{\sqrt{2}}\left(\sqrt{2}v_d + \rho_d^0(x) \right) e^{i\frac{F_d^0(x)}{\sqrt{2}v_d}},
\ea
where we have introduced the two phases $F_u$ and $F_d$ of the two neutral Higgs fields, 
information on the periodicity is  obtained by combining all the phases of $V'$
\ba
\theta(x)\equiv\frac{g_B (q_d-q_u)}{2 M}b(x)-\frac{1}{\sqrt{2}v_u} F_u^0(x) +\frac{1}{\sqrt{2}v_d} F_d^0(x).
\ea
Using the matrix $O^{\chi}$ to rotate on the physical basis of the CP-odd scalar sector, the phase describing the periodicity of the potential turns out to be proportional to the physical axion $\chi$, modulo a dimensionfull constant ($\sigma_\chi$)
\ba
\theta(x)\equiv \frac{\chi(x)}{\sigma_\chi},
\label{theta}
\ea
where we have defined
\beq
\sigma_\chi\equiv\frac{2  v_u v_d M}{\sqrt{g_B^2 (q_d-q_u)^2 v_d^2 v_u^2 +2 M^2 (v_d^2+v_u^2)}}.
\eeq
Notice that $\sigma_\chi$, in our case, takes the role of $f_a$ of the PQ case, where the angle of 
misalignment is identified by the ratio $a/f_a$, with $a$ the PQ axion. \\
As already mentioned, the re-analysis of the $V'$ potential is particularly useful for proving the gauge invariance of $\chi$ under a $U(1)_B$ infinitesimal gauge transformation with gauge parameter $\alpha_B(x)$. In this case one gets 
\beqa
\delta H_u &=& - \frac{i}{2} q_u g_B \alpha_B H_u \nonumber \\ 
\delta H_d &=& - \frac{i}{2} q_d g_B \alpha_B H_d \nonumber \\ 
\delta F_0^u &=& - \frac{v_u}{\sqrt{2}} q_u g_B \alpha_B \nonumber \\
\delta F_0^d &=& -\frac{v_d}{\sqrt{2}}  q_d g_B \alpha_B \nonumber \\
\delta b &=& -M-S \alpha_B
\eeqa
giving for \eqref{theta} $\delta\theta=0$. The gauge invariance under $U(1)_Y$ can also be easily proven using the invariance of the St\"uckelberg field $b$ under the same gauge group, sand the fact that the hypercharges of the two Higgses are equal. Finally, the invariance under $SU(2)$ is obvious since the linear combination of the phases that define $\theta(x)$ are not touched by the transformation. \\
From the Peccei-Quinn breaking potential we can extract the following periodic potential
\begin{align}
V^\prime=& 4 v_u v_d
\left(\lambda_2 v_d^2+\lambda_3 v_u^2+\lambda_0\right) \cos\left(\frac{\chi}{\sigma_\chi}\right) + 2 \lambda_1 v_u^2 v_d^2 \cos\left(2\frac{\chi}{\sigma_\chi}\right),
\label{extrap}
\end{align}
with a mass for the physical axion $\chi$  given by
\ba
m_{\chi}^2=\frac{2 v_u v_d}{\sigma^2_\chi}\left(\bar{\lambda}_0 v^2 +\lambda_2 v_d^2 +\lambda_3 v_u^2+4 \lambda_1 v_u v_d\right) 
\approx \lambda v^2.
\label{axionmass}
\ea
The size of the potential is driven by the combined product of non-perturbative effects, due to the exponentially small parameters  
$(\bar{\lambda_0}, \lambda_1,\lambda_2,\lambda_3)$, with the electroweak vevs of the two Higgses. Notice also the irrelevance of the St\"uckelberg scale $M$ in determining the value of $\sigma_\chi\sim O(v)$ and of $m_\chi$  near the transition region, due to the large suppression factor $\lambda$ in Eq.~(\ref{axionmass}). 
One point that needs to be stressed is the fact that at the electroweak epoch the angle of misalignment generated by the extra potential is parameterized by $\chi/\sigma_\chi$, while the interaction of the physical axion with the gauge fields is suppressed by $M^2/v$. This feature is obviously unusual, since in the PQ case both scales reduce to a single scale, the axion decay constant $f_a$.

\subsection{The Yukawa couplings and the axi-Higgs }

\begin{figure}[t]
\begin{center}
\includegraphics[scale=1]{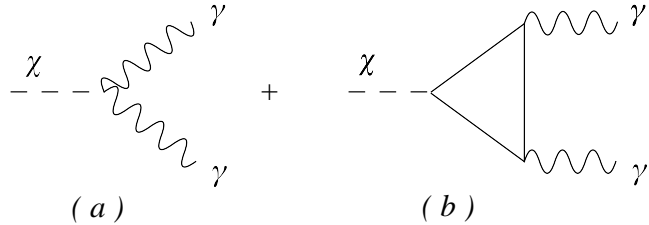}
\caption{ Contributions to the $\chi \rightarrow \g \g$ decay.\label{fig:chi_decay} describing the anomaly contribution (a) and the interaction mediated by the Yukava coupling in the fermion loop (b). }
\end{center}
\end{figure}

The Yukawa couplings determine an interaction of the axi-Higgs to the fermions. This interaction is generated by the rotation in the CP-odd sector of the scalars potential, which mixes the CP-odd components, with the inclusion of the St\"uckelberg $b$, via the matrix $O_\chi$.   
The Yukawa couplings of the model are given by
\bea
{\cal L}_{\rm Yuk}^{unit.} 
&=& - \Gamma^{d} \, \bar{Q} H_{d} d_{R} - \Gamma^{d} \, \bar{d}_R H^{\dagger}_{d} Q - 
\Gamma^{u} \, \bar{Q}_{L} (i \sigma_2 H^{*}_{u}) u_{R} 
- \Gamma^{u} \, \bar{u}_R (i \sigma_2 H^{*}_{u})^{\dagger} Q_{L} \nonumber\\
&&-   \Gamma^{e} \, \bar{L} H_{d} {e}_{R} - \Gamma^{e} \, \bar{e}_R H^{\dagger}_{d} L  \nonumber\\
&=& - \Gamma^{d} \, \bar{d} H^{0}_{d} P_{R} d - \Gamma^{d} \, \bar{d}  H^{0*}_{d} P_{L} d  
- \Gamma^{u} \, \bar{u}  H^{0*}_{u} P_{R} u - \Gamma^{u} \, \bar{u}  H^{0}_{u} P_{L} u \nonumber\\
&&- \Gamma^{e}  \, \bar{e} H^{0}_{d} P_{R} e - \Gamma^{e} \, \bar{e}  H^{0*}_{d} P_{L} e 	,
\label{yukawa_utile}
\eea
where the Yukawa coupling constants $\Gamma^{d}, \Gamma^{u}$ and $ \Gamma^{e}$  run over the three generations, i.e. $u = \{u, c, t\}$, $d = \{d, s, b\}$ and $e$ = \{$e$, $\mu$, $\tau$\}.
Rotating the CP-odd and CP-even neutral sectors into the mass eigenstates and 
expanding around the vacuum one obtains 
\begin{align}
H_u^0 =& v_u + \frac{  Re{H^0_{u}} + i \, Im{H^0_u}}{\sqrt{2}}  \nonumber\\
=&  v_u + \frac{  (h^0 \sin\a  - H^0 \cos\a ) 
+ i \, \left(O^{\chi}_{11}G^1_0 + O^{\chi}_{21}G^2_0 + O^\chi_{31} \chi	  \right) }{\sqrt{2}}  
\label{Higgs_up}   
\end{align}
\begin{align}     
H_d^0 =&  v_d + \frac{ Re{H^0_d} + i \, Im{H^0_d}}{\sqrt{2}}   \nonumber\\
=&  v_d  +  \frac{  (h^0 \cos\alpha  + H^0 \sin\alpha ) 
+ i \left( O^{\chi}_{12}G^1_0 + O^{\chi}_{22}G^1_0 + O^\chi_{32} \chi \right) }{\sqrt{2}}  
\label{Higgs_down}
\end{align}
where the vevs of the two neutral Higgs bosons $v_u=v \sin \beta $ and $v_d= v \cos \beta $ satisfy 
\ba
\tan \beta = \frac{v_u}{v_d}, \qquad  v=\sqrt{v_u^2+v_d^2}.
\ea
The fermion masses are given by
\ba
&& m_{u} =  {v_u \G^{u}},\hskip 1cm  m_{\n } =  {v_u \G^\n},  \nonumber\\
&& m_{d} =  {v_d \G^d},\hskip 1cm  m_{e} =  {v_d \G^e},
\label{f_masses}
\ea
where the generation index has been suppressed. The fermion masses, defined in terms of the two expectation values $v_u,v_d$ of the model, show an enhancement of the down-type Yukawa couplings for large values of $\tan \beta$ while at the same time the up-type Yukawa couplings get a suppression. The couplings of the $h^0$ boson to fermions are given by
\ba
{\cal L}_{\rm Yuk}(h^0) =  -  \Gamma^d \, \bar{d}_{L} d_R
\left( \frac{ \cos\a}{\sqrt{2}} h^0 \right) - \Gamma^u \, 
\bar{u}_{L} u_{R}   \left( \frac{ \sin\a }{\sqrt{2}} h^0  \right)  
-  \Gamma^e \, \bar{e}_{L}e_R  \left( \frac{ \cos\a}{\sqrt{2}} h^0 \right)  + c.c. 
\ea
The couplings of the $H^0$ boson to the fermions are 
\ba
{\cal L}_{\rm Yuk}(H^0) =  - \Gamma^d \, \bar{d}_{L} d_R 
\left( \frac{ \sin\a}{\sqrt{2}} H^0 \right) -   \Gamma^u \, 
\bar{u}_{L} u_{R}   \left( - \frac{ \cos\a }{\sqrt{2}} H^0  \right)
 - \Gamma^e \, \bar{e}_{L}e_R  \left( \frac{ \sin\a}{\sqrt{2}} H^0 \right) + c.c. 
\ea
The interaction of $\chi$ with the fermions is proportional to the rotation matrix $O^{\chi}$ and to the mass of the fermion.
The decay of the axi-Higgs is driven by two contributions, the direct point-like WZ interaction  $(\chi/M F \tilde{F})$ and the fermion loop. The amplitude can be separated in the form 
corresponding to the two contributions from diagrams a) and b) of Fig.~\ref{fig:chi_decay} 
\ba
{\mathcal M}^{\mu \nu}(\chi \rightarrow \g \g) = {\mathcal M}^{\mu \nu}_{WZ}+{\mathcal M}^{\mu \nu}_{f}.
\ea
The direct coupling related to the anomaly is given by the vertex
shown in Fig.~\ref{fig:chi_decay} a)
\bea
{\mathcal M}^{\mu \nu}_{WZ}(\chi \rightarrow \g \g) = 4 g^{\chi}_{\g\g} \varepsilon[\mu,\nu,k_1,k_2]
\eea
coming from the WZ counterterm $\chi F^{}_\g \tilde{F}^{}_\g$ which gives a decay rate of the form 
\ba
\label{WZrate1}
\Gamma^{}_{WZ}(\chi \rightarrow \g\g)= \frac{m^3_\chi}{4 \pi}(g^\chi_{\g\g})^2.
\ea
We remark that $g^{\chi}_{\g \g}$ is of $O(g^3 v/M^2)$, as derived from Eq. (\ref{gchi}), with charges that have been chosen of $O(1)$. \\
It is 

\begin{figure}[t]
\begin{center}
\includegraphics[scale=0.3, angle=-90]{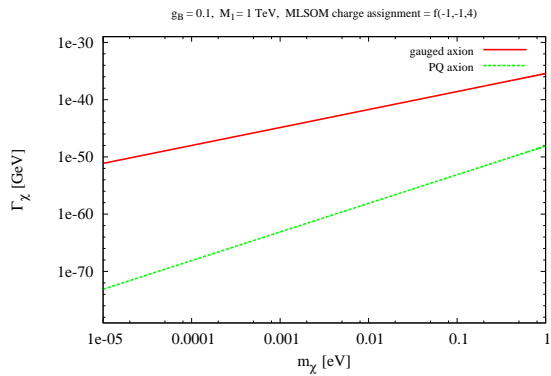}
\end{center}
\caption{\small Total decay rate of the axi-Higgs for several mass values.
Here, for the PQ axion, we have chosen $f_a=10^{10}$ GeV.}
\label{decs1}
\end{figure}

Comparative studies of the decay rate into photons for the axi-Higgs with the ordinary PQ axion have been performed for a St\"uckelberg scale confined in the TeV range 
and a mass of $\chi$ in the same range expected for the PQ axion. The analysis shows that the total decay rate of $\chi$ into photons is of the order $\Gamma_\chi\sim 10^{-50}$ GeV, 
which is larger than the decay rate of the PQ axion in the same channel $(10^{-60})$, but small enough to be long- lived, with a lifetime larger than the age of the universe. We show in Fig. \ref{decs1} the result of this study, where we compare predictions for the decay rate of the axi-Higgs into two photons to that of the ordinary PQ axion. \\
The charge assignment of the anomalous model have been denoted as $f(-1,1,4)$, where we have used the convention
\bea
f(q_{Q_L}^B, q_{L}^B, \Delta q^B)\equiv
(q_{Q_L}^B, q_{u_R}^B; q^{B}_{d_R}, q^{B}_{L}, q^{B}_{e_R}, q^{B}_{u},q^{B}_{d}).
\eea
These depend only upon the three free parameters $q^B_{Q_L}$, $q^B_{L},\Delta q^B$. The parametric solution of the anomaly equations of the model $f(q^B_{Q_L}, q^B_{L}, \Delta q^B)$,
for the particular choice $q_{Q_L}^B=-1, q_{L}^B=-1$, reproduces the entire charge assignment of a special class of intersecting brane models (see \cite{Ibanez:2001nd} and \cite{Ghilencea:2002da} and the discussion in \cite{Coriano:2009zh})
\bea
f(-1, -1, 4)=(-1, 0, 0, -1, 0, +2, -2).
\eea
We refer to \cite{Coriano:2010py} for further details on these studies.


\section{Relic density for a low ($\sim$ 1 TeV) St\"uckelberg scale } 
The computation of the relic density for the St\"uckelberg axi-Higgs can be performed as in \cite{Coriano:2010ws}, adopting a low scale scenario, where the extra $V'$ \eqref{extrap} potential which causes the vacuum misalignment is generated around the electroweak scale. 

\begin{figure}[t]
\centering
\includegraphics[scale=.5]{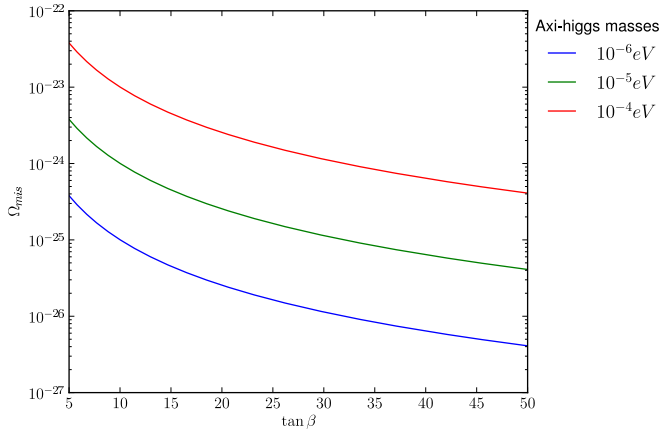}
\caption{Relic density of the axi-Higgs as a function of $\tan\beta$ for several values of the mass of the axi-Higgs.}
\label{fig:relicvu}
\end{figure}

One starts from the Lagrangian
\begin{align} 
\mathcal{S}=
\int d^4 x \sqrt{g}\left( \frac{1}{2}\dot{\chi}^2 - \frac{1}{2}
m_\chi^2 \Gamma_\chi \dot{\chi}\right), 
\end{align} 
where $\Gamma_\chi$ is
the decay rate of the axion, where the potential has been expanded around
its minimum up to quadratic terms. The same action can be derived from the
quadratic approximation to the general expression 
\begin{align}
\mathcal{S}=\int d^4 x R^3(t)\left( \frac{1}{2} \sigma_{\chi}^2
\left(\partial_{\alpha} \theta\right)^2 - \mu^4 \left(1-
\cos\theta\right) - V_0\right) 
\end{align} 
which, as just mentioned, is constructed
from the expression of $V^\prime$ given in Eq.~(\ref{extrap}). Here
$\mu\sim v$, is the electroweak scale. We also set to zero other contributions to
the vacuum potential ($V_0=0$). In a 
Friedmann-Robertson-Walker spacetime metric, with a scaling factor
$R(t)$, this action gives the equation of motion 
\begin{align}
\frac{d}{dt}\biggl[\left( R^3(t) (\dot{\chi} +
\Gamma_\chi\right)\biggr] + R^3 m_{\chi}^2(T) =0.
\label{FRWequation}  
\end{align}
We will neglect the decay rate of the axion
in this case and set $\Gamma_\chi\approx 0$. At this point, we are free to set the scale at which the $V^\prime$ potential, which is of non-perturbative origin, is generated. Therefore it will be zero above the electroweak scale (or temperature
$T_{ew}$), which will give $m_\chi=0$ for $T\gg T_{ew}$. The general equation
of motion derived from Eq.~(\ref{FRWequation}), introducing a
temperature dependent mass, can be written as
\begin{align} 
\ddot{\chi} + 3 H \dot{\chi} + m_{\chi}^2(T) \chi =0,
\label{motioneq}
\end{align} 
which allows as a solution a constant value of
the misalignment angle $\theta=\theta_{i}$. The axion energy density is given by 
\begin{align}
\rho=\frac{1}{2}
\dot{\chi}^2 + \frac{1}{2} m_\chi^2 \chi^2,
\label{rhoeq}  
\end{align}
which after a harmonic averaging, due to the periodic motion, gives 
\begin{align}
\langle
\rho\rangle = m_\chi^2 \langle \chi^2\rangle.
\label{averageeq}  
\end{align}
By differentiating Eq.~(\ref{rhoeq}) and using 
the equation of motion in~(\ref{motioneq}), followed by the 
averaging Eq.~(\ref{averageeq}) one obtains the relation
\begin{align} 
\langle \dot{\rho}\rangle =\langle \rho\rangle \left(
-3 H + \frac{\dot{m}}{m}\right),
\end{align} 
with a mass which is time-dependent through its
temperature $T(t)$, while $H(t)=\dot{R}(t)/R(t)$ is the Hubble
parameter. One easily finds that the solution of this
equation is of the form
\begin{align} 
\langle \rho\rangle = \frac{m_\chi(T)}{R^3(t)}
\end{align} 
which shows the decay of the energy density with an
increasing space volume, valid even for a $T$-dependent mass. The condition for the oscillations of $\chi$ to take place is that the  the universe has to be old at least as the the period of oscillation. Then the axion field starts oscillating and appears as dark matter, otherwise $\theta$ is misaligned but frozen. This is the physical content of the condition
\begin{align} 
m_\chi(T_{i})= 3 H(T_{i}),
\label{mhcond}
\end{align} which allows to identify the initial temperature of the
coherent oscillation of the axion field $\chi$, $T_i$, by equating
$m_\chi(T)$ to the Hubble rate, taken as a function of temperature.\\
In the radiation era, the thermodynamics of all the components of the
primordial state is entirely determined by the temperature $T$, being
the system at equilibrium. This is because the contents of the early universe were in approximate  thermal equilibrium, being the interaction rates of the constituents were large compared to the interaction rates $H$.\\
Pressure and entropy are then just
given as a function of the temperature
\begin{align} & \rho=3 p=\frac{\pi^2}{30} g_{*,T}T^4 \nonumber \\ &
s=\frac{2 \pi^2}{45} g_{*,S,T} T^3.
\label{entropy}
\end{align} Combined with the Friedmann equation they allow to relate the Hubble
parameter and the energy density
\begin{equation} H=\sqrt{\frac{8}{3} \pi G_N \rho},
\label{hubble}
\end{equation} with $G_N={1}/{M_P^2}$ being the Newton constant and
$M_P$ the Planck mass. The number density of axions $n_\chi$ decreases
as $1/R^3$ with the expansion, as does the entropy density $s\equiv
S/R^3$, where $S$ indicates the comoving entropy density, which
remains constant in time, leaving the ratio $Y_a\equiv
n_\chi/s$ conserved. An important variable is the abundance of
$\chi$ at the temperatire of oscillations $T_i$, which is defined as
\begin{equation} Y_\chi(T_i)= \frac{n_\chi}{s}\bigg\vert_{T_{i}}.
\end{equation}. At the beginning of the oscillations the total energy density is
just the potential one
\begin{equation} \rho_\chi=n_\chi(T_{i}) m_\chi(T_{i})=1/2
m_\chi^2(T_i)\chi_i^2,
\end{equation} giving for the initial abundance at $T=T_{i}$
\begin{equation} Y_\chi(T_{i})= \frac{1}{2}\frac{m_\chi(T_{i})
\chi_i^2}{s}= \frac{45 m_\chi(T_{i})\chi_i^2}{4 \pi^2 g_{*,S,T} T_{i}^3}
\label{yeq}
\end{equation} where we have used the expression
of the entropy  given by
Eq.~(\ref{entropy}).  At this point, by inserting the expression of $\rho$
given in Eq.~(\ref{entropy}) into the expression of the Hubble rate as
a function of density given by Eq.~(\ref{hubble}), the condition for
oscillation Eq.~(\ref{mhcond}) allows to express the axion mass at
$T=T_{i}$ in terms of the effective massless degrees of freedom
evaluated at the same temperature
\begin{equation} m_\chi(T_{i})=\sqrt{\frac{4}{5}\pi^3
g_{*,T_{i}}}\frac{T_i^2}{M_P}.
\label{Tmass}
\end{equation} This gives for Eq.~(\ref{yeq}) the expression
\begin{equation} Y_\chi(T_{i})= \frac{45
\sigma_\chi^2\theta_i^2}{2\sqrt{5 \pi g_{*, T_{i}}} T_{i} M_P},
\label{ychi}
\end{equation} where we have expressed $\chi$ in terms of the angle of
misalignment $\theta_i$ at the temperature when oscillations start. Notice that we are
assuming that $\theta_i=\langle \theta\rangle$ is the zero mode of the
initial misalignment angle after an averaging. \\
$g_{*,T}=110.75$ is the number of massless degrees of freedom of the model at the electroweak scale.
Using the conservation of the abundance $Y_{a 0}=Y_{a}(T_{i})$, the expression of the contribution to the relic density  is given by 
\begin{equation}
\Omega_\chi^{mis}=\frac{n_\chi}{s}\bigg\vert_{T_{i}} m_\chi\frac{s_0}{\rho_c}.
\label{omegaeq}
\end{equation}
To evaluate \eqref{omegaeq} we need the values of the critical energy density ($\rho_c$) and the entropy density today, which are estimated as
\begin{equation}
\rho_{c}=5.2\cdot10^{-6}\textrm{GeV}/\textrm{cm}^3\hspace{1cm}s_0=2970 \,\,\textrm{cm}^{-3},
\end{equation}
with $\theta\simeq1$. 
Given these values, the relic density as a function of $\tan\beta=v_u/v_d$, the ratio of the two Higgs vevs, is given in Fig. \ref{fig:relicvu}. In this plot we have varied the oscillation mass and plotted the relic densities as a function of this variable.  The variation of $v_u$ has been constrained  to give the values of the masses of the electroweak gauge bosons, via an appropriate choice of $\tan\beta$. \\
For instance, if we assume a temperature of oscillation 
of $T_i=100$ \textrm{GeV}, an upper bound for the axi-Higgs mass, which allows the oscillations to take place, is $m_\chi(T_i)\approx 10^{-5} \textrm{eV}$, with $g_{*,T} \approx 100$. \\
In order to specify $\sigma_{\chi}$ we have assumed a value of 1 TeV for the St\"uckelberg mass $M_S$, with a gauge coupling of the anomalous $B_\mu$, $g_B\approx1$, and we have taken $(q_u,q_d)$ of order unity, obtaining $\sigma_{\chi}\simeq \,10^2$\,GeV.
As we lower the oscillation temperature (and hence the mass), the corresponding curves for $\Omega_\chi$ are down-shifted. \\
The plot shows that the values of these relic densities at current time are basically vanishing and these small results are to be attributed to the value of $\sigma_\chi$, which is bound to vary around the electroweak scale. 
We remind that in the PQ case $\sigma_\chi$ is replaced by the large scale $f_a$ at the QCD phase transition, which determines an enhancement of $\Omega_\chi$ respect to the current case.\\
As already mentioned, nonperturbative instanton effects at the electroweak scale are expected to vastly suppress the mass of the axi-Higgs, as derived in \eqref{axionmass}, in the form 
\begin{equation}
m_\chi^2\sim\Lambda_{ew}^4/v^2, \qquad \textrm{with} \qquad \Lambda_{ew}^4\sim {\textrm {Exp}}(-2\pi/\alpha_w(v)) v^4 
\end{equation}
$ \alpha_W(v)$ being the weak charge at the scale $v$ - which is indeed a rather small value since ${\textrm{ Exp}}(-2\pi/\alpha_w(v))\sim e^{-198}$. We will come back to this point in the next section, when discussing the possibility of raising $M_S$ from the TeV range up to the GUT or Planck scales.\\
For this reason $\chi$ remains essentially a physical but frozen degree of freedom which may undergo a significant (second) misalignment only at the QCD phase transition. The possibility of sequential misalignments has been taken into account both in non supersymmetric \cite{Coriano:2010py} and in supersymmetric models \cite{Coriano:2010ws}.
It is the presence of a coupling of the axion to the gluons, via the color/ $U(1)_B$ mixed anomaly, that $\chi$ behaves, in this case, similar to a PQ axion. The misalignment is controlled by the periodic potential generated at the QCD phase transition, being the first misalignment at the electroweak scale irrelevant. In the absence of such mixed anomaly, $\chi$ could be classified as a quintessence axion, contributing to the dark energy content of the universe.

\begin{figure}[t]
\centering
\includegraphics[scale=.5]{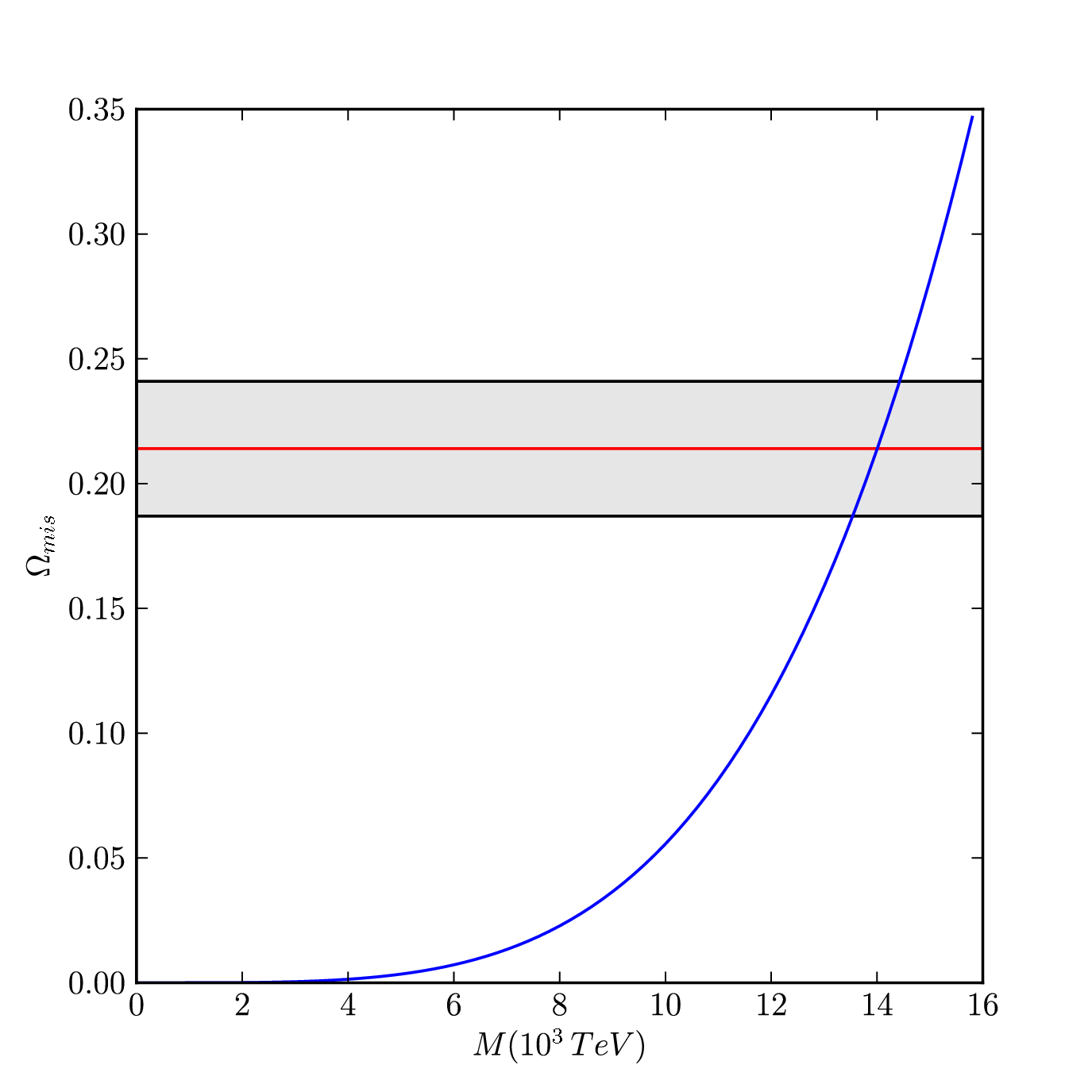}
\caption{Relic density of the axi-Higgs as a function of $M$. The grey bar represents the measured value of $\Omega_{DM} h^2=0.1123\pm0.0035$\label{fig:relicM}}
\end{figure}

We show in Fig. \ref{fig:relicM} results of a numerical study of $\Omega_{mis}h^2$ as a function of $M_S$, expressed in units of $10^9$ GeV. We show as a darkened area the bound coming from WMAP data~\cite{Jarosik:2010iu}, given as the average value plus an error band,  while the monotonic curve denotes the values of $\Omega_{mis}h^2$ as a function of $M_S$. It is clear that the relic density of $\chi$ can contribute significantly to the dark matter content only if the St\"uckelberg scale is rather large ($\sim 10^7$ GeV) and negligible otherwise. \\
In the next section we are going to address another scenario, where we will assume that the St\"uckelberg scale is around the Planck scale and the breaking of the symmetry which 
allows to generate a periodic potential for the $b$ fiels is taken at the GUT scale. This particular choice for the location of the two scales, which is well motivated in a string/brane theory context, 
opens up the possibility of having an ultra-light axion in the spectrum. The De Broglie wavelength of this hypothetical particle would be around 10 kpc, which is what is required to solve the 
issues in the modelling of the matter distribution at the sub-galactic scale, that we have discussed in the introduction.

\section{ St\"uckelberg models at the Planck/GUT scale and fuzzy dark matter} 

By raising the St\"uckelberg mass near the Planck scale, the St\"uckelberg construction acquires a fundamental meaning since it can be directly related to the cancellation of a gauge anomaly 
generated at the same scale \cite{Coriano:2017ghp}. As mentioned above, anomalous $U(1)$ symmetries are quite generally present in theories of intersecting branes.
However, the very same structure emerges also in the low energy limit of heterotic string constructions.
At the same time, as shown in \cite{Coriano:2005js}, even in the presence of multiple anomalous abelian symmetries, only a single axion is necessary to cancel all anomalies, giving a special status to the St\"uckelberg field.
These considerations define a new context in which to harbour such models. 
In this context, it is natural to try to identify a consistent formulation within an ordinary gauge theory, by assuming that the axion emerges at the Planck scale $M_P$, but it acquires a mass at a scale below, 
which in our case is assumed to be the GUT scale. 
In this section therefore we are going to consider an extension of the setup discussed in previous sections, under the assumption that their dynamics is now controlled by two scales.

We will consider an $E_6$ based model, derived from $E_8$, which appeared in the heterotic string construction of \cite{Gross:1984dd} with an
$E(8) \times  E(8)$ symmetry. After a compactification of six spatial dimensions on a Calabi-Yau manifold \cite{Candelas:1985en} the symmetry is reduced to an $E(6)$ GUT gauge theory. 
Other string theory compactifications predict different GUT gauge structures, such as  $SU(5)$ and $SO(10)$. The $E_6$, however, allows to realize a scenario where two components of dark matter are present, as we are going to elaborate.
Fermions are assigned to the ${\bf 27}$ representation of $E(6)$, which is anomaly-free.
Notice that in $E(6)$ a PQ symmetry is naturally present, as shown in \cite{Frampton:1981ik}, which allows to have an ordinary PQ axion, while at the same time it is a realistic GUT symmetry which can break to the SM. This is the gauge structure to which one may append an anomalous $U(1)_X$ symmetry.\\
We consider a gauge symmetry of the form $E_6\times U(1)_X$, where the gauge boson $B^\mu$ is in the St\"uckelberg phase. $B_{\alpha}$ is the gauge field of $U(1)_X$
and $B_{\alpha\beta} \equiv \partial_{\alpha}B_{\beta} - \partial_{\beta}B_{\alpha}$  the corresponding field strength, while $g_B$ its gauge coupling. As already mentioned, the $U(1)_X$ carries an anomalous coupling to the fermion spectrum. 

The one-particle irreducible (1PI) effective Lagrangian of the theory at 1-loop level takes the form 
\begin{equation}
{\cal L}={\cal L}_{E_6}+ {\cal L}_{St} + {\cal L}_{anom} + {\cal L}_{WZ},
\end{equation}
in terms of the gauge contribution of $E_6$ (${\cal L}_{E_6}$), the St\"uckelberg term ${\cal L}_{St}$, the anomalous 3-point functions  ${\cal L}_{anom}$, generated by the anomalous fermion couplings to the $U(1)_X$ gauge boson, and the Wess-Zumino counterterm (WZ) ${\cal L}_{WZ}$. 
The St\"uckelberg interaction to the $E_6$ gauge Lagrangian
\begin{equation}
{\cal L}_{E_6}=-\frac{1}{4} F^{(E_6)\, \mu\nu}F^{(E_6)}_ {\mu\nu}, 
\end{equation}
which enables us to write the St\"uckelberg part of the lagrangian as
\begin{equation}
{\cal L}_{Stueck} = - \frac{1}{4} B_{\alpha\beta} B^{\alpha\beta} - \frac{1}{2} (M B_{\alpha} - \partial_{\alpha} b(x))^2.
\end{equation}
\noindent
In this final form, $M$ is the mass of the St\"uckelberg gauge boson associated with $U(1)_X$ which we can be taken of the order of the Planck scale, guaranteeing the decoupling of the axion around $M_{GUT}$, due to the gravitational suppression of the WZ counterterms.  
The WZ contribution is the combination of two terms 
\begin{equation}
{\cal L}_{WZ}= c_1 \frac{b}{M}F^{(E_6)\, \mu\nu}F^{(E_6)\, \rho\sigma}\epsilon_{\mu\nu\rho\sigma} + 
c_2  \frac{b}{M}B_{\mu\nu}B_{\rho\sigma}\epsilon^{\mu\nu\rho\sigma} 
\label{WZ}
\end{equation}
needed for the cancellation of the $U(1)_X E_6 E_6$  and $U(1)_X^3$ anomalies, for appropriate values of the numerical constants $c_1$ and $c_2$, fixed by the charge assignments of the model.
The three chiral familes will be assigned under $E(6)\times U(1)_X$ respectively to
\begin{equation}
{\bf 27}_{X_1}  ~~~~~ {\bf 27}_{X_2} ~~~~~ {\bf 27}_{X_3},
\label{chiral}
\end{equation}
in which the charges $X_i$ ($i=1,2,3$) are free at the moment, while the cancellation of 
the $U(1)_X^3$ and $E_6\times U(1)_X^2$ anomalies implies that
\begin{equation}
\sum_{i=1}^{3} X_i^3 = 0,     \qquad\qquad \sum_{i=1}^3 X_i=0.
\label{Xcharges}
\end{equation}
These need to be violated in order to compensate with a Wess-Zumino term for the restoration of the gauge symmetry of the action.\\
Concerning the scalar sector, this contains two ${\bf 351}_{X_i}$ $(i=1,2)$ irreducible representations, where the $U(1)_X$ charges $X_i$ need to be determined. 
The ${\bf 351}$ is the {\it antisymmetric} part of the Kronecker product ${\bf 27} \otimes {\bf 27}$ where ${\bf 27}$ is the defining representation of $E(6)$. The ${\bf 351}_X$ can be conveniently described by the 2-form $A_{\mu\nu} = - A_{\nu\mu}$ with $\mu,\nu = 1$ to $27$. The 
most general renormalizable potential in ${\cal L}_{E_6}$ is expressed in terms of $A^{(1)}_{\mu\nu}$ and $A^{(2)}_{\mu\nu}$ of $U(1)_X$ of charges $x_1$ and $x_2$ respectively. 
If we denote the ${\bf 27}_{X_i}$ of Eq.(\ref{chiral}) by $\Psi_{\mu}$ with $\mu = 1$ to $27$ then the full Lagrangian including the potential $V$, has an invariance under the global symmetry 
 \begin{equation}
 A^{(1)}_{\mu\nu} \rightarrow e^{i\theta} A^{(1)}_{\mu\nu}~~~~~A^{(2)}_{\mu\nu} \rightarrow e^{i\theta} A^{(2)}_{\mu\nu} ~~~~ \Psi_{\mu} \rightarrow e^{- (\frac{1}{2}i\theta)} \Psi_{\mu}.
 \label{PQ}
 \end{equation}
This is identifiable as a Peccei-Quinn symmetry which is broken at the GUT scale when $E(6)$
 is broken to $SU(5)$ \cite{Frampton:1981ik}. This axionic symmetry can be held responsible for solving the strong CP problem. 
  \noindent
 We couple $A^{(1)}_{\mu\nu}$ to the fermion families $({\bf 27})_{X_i}$ $i=1,2,3$. We choose
 in Eq. (\ref{chiral}), {\it e.g.} $X_1=X_2=X_3=+1$, with the $X$-charge of $A^{(1)}$ fixed to $X=-2$. The second scalar representation $A^{(2)}$ is decoupled from the fermions, with an 
 $X-$charge for $A^{(2)}$ which is arbitrary and taken for simplicity to be $X=+2$. 
 The potential is expressed in terms of three $E_6\times U(1)_X$ invariant components, 
\begin{equation}
V= V_1+  V_2 + V_p,
\end{equation}
where
\begin{equation} 
V_1= F(A^{(1)},A^{(1)})  \qquad  V_2= F(A^{(2)},A^{(2)}),
\end{equation}
with $V_1$ and $V_2$ denoting the contributions of $({\bf 351})_{-2}$ and $({\bf 351})_{+2}$, expressed in terms of 
the function \cite{Frampton:1981ik}
\begin{eqnarray}
 F(A^{(i)},A^{(j)}) & = &\left. M_{GUT}^2 A^{(i)}_{\mu\nu} \bar{A^{(j)}}^{\mu\nu} +h_1 ~(A^{(i)}_{\mu\nu} \bar{A^{(j)}}^{\mu\nu})^2  +h_2  ~ A^{(i)}_{\mu\nu} \bar{A}^{\nu\sigma} A^{(i)}_{\sigma\tau}\bar{A}^{\tau\mu} \right.\nonumber \\
&&\left.\qquad \qquad +\,h_3 ~  d^{\mu\nu \lambda} d_{\xi\eta\lambda} A^{(i)}_{\mu\sigma}A^{(i)}_{\nu\tau} \bar{A^{(j)}}^{\xi\sigma} \bar{A^{(j)}}^{\eta\tau} \right. \nonumber \\
 & &\left.  \qquad \qquad +\, h_4 ~ d^{\mu\nu\alpha}d^{\sigma\tau\beta}d_{\xi\eta\alpha} d_{\lambda\rho\beta} A^{(i)}_{\mu\sigma}A^{(i)}_{\nu\tau} \bar{A^{(j)}}^{\xi\lambda} \bar{A^{(j)}}^{\eta\rho}\right. \nonumber \\
 & &\left. \qquad \qquad + \,h_5 ~ d^{\mu\nu\alpha} d^{\sigma\beta\gamma} d_{\xi\eta\beta} d_{\lambda\alpha\gamma} A^{(i)}_{\mu\sigma}A^{(i)}_{\nu\tau} \bar{A^{(j)}}^{\xi\lambda} \bar{A^{(j)}}^{\eta\tau} \right.\nonumber \\
 & &\left. \qquad \qquad +\,h_6 ~ d^{\mu\nu\alpha}d^{\sigma\tau\beta} d_{\alpha\beta\gamma} d^{\gamma\zeta\xi} d_{\xi\eta\zeta} d_{\lambda\rho\chi} A^{(i)}_{\mu\sigma}\bar{A^{(j)}}^{\xi\lambda} A^{(i)}_{\nu\tau} \bar{A^{(j)}}^{\eta\rho}\right. ,
 \end{eqnarray}
 in which $d_{\alpha\beta\gamma}$ with $\alpha,\beta,\gamma = 1$ to $27$ is the $E(6)$
invariant tensor.\\
As for the two Higgs doublet model discussed in the previous sections, also in this case we are allowed to introduce a periodic potential on the basis of the underlying gauge symmetry, of the form
\begin{eqnarray}
V_p & = & M_{GUT}^2 A^{(1)}_{\mu\nu} \bar{A^{(2)}}^{\mu\nu}e^{- i 4\frac{b}{M_S}}   + e^{- i 8 \frac{b}{M_S}}\left[(h_1 ~(A^{(1)}_{\mu\nu} \bar{A^{(2)}}^{\mu\nu} )^2  + h_2  ~ A^{(1)}_{\mu\nu} \bar{A^{(2)}}^{\nu\sigma} A^{(1)}_{\sigma\tau}\bar{A^{(2)}}^{\tau\mu} \right.\nonumber \\
&&\left.\qquad \qquad +\,h_3 ~  d^{\mu\nu \lambda} d_{\xi\eta\lambda} A^{(1)}_{\mu\sigma}A^{(1)}_{\nu\tau} \bar{A^{(2)}}^{\xi\sigma} \bar{A^{(2)}}^{\eta\tau} \right. \nonumber \\
 & &\left.  \qquad \qquad +\, h_4 ~ d^{\mu\nu\alpha}d^{\sigma\tau\beta}d_{\xi\eta\alpha} d_{\lambda\rho\beta} A^{(1)}_{\mu\sigma}A^{(1)}_{\nu\tau} \bar{A^{(2)}}^{\xi\lambda} \bar{A^{(2)}}^{\eta\rho}\right. \nonumber \\
 & &\left. \qquad \qquad + \,h_5 ~ d^{\mu\nu\alpha} d^{\sigma\beta\gamma} d_{\xi\eta\beta} d_{\lambda\alpha\gamma} A^{(1)}_{\mu\sigma}A^{(1)}_{\nu\tau} \bar{A^{(2)}}^{\xi\lambda} \bar{A^{(2)}}^{\eta\tau} \right.\nonumber \\
 & &\left. \qquad \qquad +\,h_6 ~ d^{\mu\nu\alpha}d^{\sigma\tau\beta} d_{\alpha\beta\gamma} d^{\gamma\zeta\xi} d_{\xi\eta\zeta} d_{\lambda\rho\chi} A^{(1)}_{\mu\sigma}\bar{A^{(2)}}^{\xi\lambda} A^{(1)}_{\nu\tau} \bar{A^{(2)}}^{\eta\rho}\right] + h.c.
 \end{eqnarray}
and which becomes periodic at the GUT scale after symmetry breaking, similarly to the case considered in 
\cite{Coriano:2010py,Coriano:2010ws}. This potential is expected to be of nonperturbative origin and generated at the scale of the GUT phase transition. Also in this case the size of the 
 contributions in $V_p$, generated by instanton effects at the GUT scale, are expected to be exponentially suppressed. However, the size of the suppression is related to the value of the gauge coupling at the corresponding scale.  

 \subsection{The periodic potential}
 
The breaking of the $E_6\times U(1)_X$ symmetry at $M_{GUT}$ can follow different routes such as  
  $E(6) \supset SU(3)_C \times SU(3)_L \times SU(3)_H$ where 
\begin{eqnarray}
\label{351prime}
({\bf 351}) & = & (1, 3^*, 3) + (1, 3^*,6^*) + (1, 6, 3) + (3, 3, 1) + (3, 6^*, 1) + (3, 3, 8) + \nonumber \\
& & (3^*, 1, 3^*) + (3^*, 1, 6)
+ (3^*, 8, 3^*) + (6^*, 3, 1) + (6, 1, 3^*) + (8, 3^*, 3)
\label{351}
\end{eqnarray}
of which the colour singlets are only the 45 states for each of the two $({\bf 351})_{X_i}$
\begin{equation}
(1, 3^*, 3)_{X_i} \qquad   (1, 3^*, 6^*)_{X_i} \qquad (1, 6, 3)_{X_I}  ~~~  i=1,2.
\label{colorsinglets}
\end{equation}
One easily realizes that there are exactly nine colour-singlet $SU(2)_L$-doublets in the $({\bf 351}^{'})_{-2}$ and 9 in the 
$({\bf 351}^{'})_{+2}$, that we may denote as  $H^{(1)}_j$, $H^{(2)}_j$, with $j=1,2\ldots 9$, which appear in the periodic potential in the form 
\begin{eqnarray}
V_p &\sim&\sum_{j=1}^{12}\lambda_0 M_{\textrm {GUT}}^2 (H^{(1)\dagger }_j H^{(2)}_j e^{- 4 i g_B\frac{b}{M_S}})+
\sum_{j,k=1}^{12}\left[\lambda_1(H^{(1)\dagger}_j H^{(2)}_j e^{-i 4 g_B \frac{b}{ M_S}})^2+\lambda_2(H_i^{(1)\dagger}H_i)(H_i^{(1)\dagger}H^{(2)}_j e^{-i 4 g_B\frac{b}{M_S}})\right.\nonumber \\
&&\left. + \lambda_3(H_k^{(2)\dagger}H_k^{(2)})(H_j^{(1)\dagger}H_k^{(2)} e^{-i 4 g_B\frac{b}{M_S}}) \right] +\textrm{h.c.},
\end{eqnarray}
where we are neglecting all the other terms generated from the decomposition (\ref{351prime}) which will not contribute to the breaking. The assumption that such a potential is instanton generated at the GUT scale, 
with parameters $\lambda_i$'s induces a specifc value of the instanton suppression which is drastically different from the case of a St\"uckelberg scale located at TeV/multi TeV range. 

For simplicity we will consider only a typical term in the expression above, involving two neutral components, generically denoted as $H^{(1)\, 0}$ and $H^{(2)\, 0}$, all the remaining contributions being similar. 
In this simplified case the axi-Higgs $\chi$ is generated by the mixing of the CP odd components of two neutral Higgses. The analysis follows rather closely the approach discussed before, in the simplest two-Higgs doublet model, which defines the template for such constructions.\\
Therefore, generalizing this procedure, the structure of $V_{p}$ after the breaking of the $E_6\times U(1)_X$ symmetry can be summarised in the form 
\begin{align}
V_p\sim&  v_1 v_2
\left(\lambda_2 v_2^2+\lambda_3 v_1^2+\overline{\lambda_0} M_{GUT}^2\right) \cos\left(\frac{\chi}{\sigma_\chi}\right) +  \lambda_1 v_1^2 v_2^2 \cos\left(2\frac{\chi}{\sigma_\chi}\right),
\label{extrap1}
\end{align}
with a mass for the physical axion $\chi$  given by
\begin{equation}
m_{\chi}^2\sim\frac{2 v_1 v_2}{\sigma^2_\chi}\left(\bar{\lambda}_0 v_1^2 +\lambda_2 v_2^2 +\lambda_3 v_1^2+4 \lambda_1 v_1 v_2\right) 
\approx \lambda v^2
\label{axionmass}
\end{equation}
with $v_1\sim v_2\sim v\sim M_{GUT}$. Assuming that $M_S$, the St\"uckelberg 
mass, is of the order of $M_{\textrm{Planck}}$ and that the breaking of the $E_6\times U(1)_X$ symmetry takes place at the GUT scale
$M_{GUT}\sim 10^{15}$ GeV,  (e.g. $v_1\sim v_2\sim M_{GUT}$) then 
\begin{equation}
\sigma_{\chi}\sim M_{\textrm{GUT}} + {\cal O}(M_{\textrm{GUT}}^2/M_{\textrm{Planck}}^2),  \qquad m_\chi^2\sim \lambda_0 M_{\textrm{GUT}}^2,
\end{equation}
where all the $\lambda_i$'s in $V_p$ are of the same order.
The potential $V_p$ being generated by the instanton sector, the size of the numerical coefficients appearing in its expression are constrained to specific values. 
One obtains $\lambda_0\sim e^{-2 \pi/\alpha(M_{\textrm{GUT}})}$, with the value of the coupling  $4 \pi g_B^2 =\alpha_{\textrm{GUT}}$ fixed at the GUT scale. If we assume that  $1/33 \le \alpha_{GUT} \le 1/32$, then  $e^{-201}\sim 10^{-91}\le \lambda_0 \le e^{-205}\sim 10^{-88}$, and
the mass of the axion $\chi$ takes the approximate value 
\begin{equation}
10^{-22}   \textrm{ eV} < m_{\chi} <  10^{-20} \textrm{ eV}, 
\end{equation}
which contains the allowed mass range for an ultralight axion, as discussed in recent analysis of the astrophysical constraints on this type of dark matter \cite{Hui:2016ltb}.\\
\subsection{Detecting ultralight axions} 
One of the interesting issues on which future research has to concentrate concerns  the possibility of suggesting new ways for detecting such specific class of particles. Several proposals for the detection of generic ultralight bosons \cite{Dev:2016hxv,Brito:2017zvb,Baumann:2018vus} in the astrophysical context have been recently presented. For instance, it has been observed  that light boson fields around spinning black holes can trigger superradiant instabilities, which can be strong enough to imprint gravitational wave detection. This could be used to set constraints on their masses and couplings. Other proposals \cite{Fukuda:2018omk} have suggested to use the precise astronomical ephemeris as a way to detect such a light dark matter, as celestial solar system bodies feel the dark matter wind which acts as a resistant force opposing their motions. The bodies feel the dark matter wind because our solar system moves with respect to the rest frame of the dark matter halo, so that the scattering off the dark matter acts as a resistant force opposing their motions.\\
It is at the moment an open issue, from our perspective, how to distinguish between the various proposals that have been put forward in the recent literature. The models that we have presented are, however, very specific, since they are accompanied by a well defined gauge structure and are, as such, susceptible of in depth analysis. 
We should also mention that another specific property of such models is their interplay with the flavour sector, especially the neutrino sector, together with their impact on leptogenesis and $SO(10)$ grand unification. This would allow to establish a possible link between the neutrino mass spectrum and the axion mass and would be an intermediate step to cover prior to a discussion of the general astrophysical suggestions for their detections mentioned above. An in-depth analysis of some of these issues is underway.

\section{Conclusions} 

The invisible axion owes its origin to a global $U(1)_{PQ}$  (Peccei-Quinn, PQ) symmetry which is spontaneously broken in the early universe and explicitly broken to a discrete $Z_N$ symmetry by instanton effects at the QCD phase transition \cite{Sikivie:1982qv}. 
The breaking occurs at a temperature $T_{PQ}$ below which the symmetry is nonlinearly realized. 
Two distinctive features of an axion solution - as derived from the original Peccei-Quinn (PQ) proposal \cite{Peccei:1977ur} and its extensions 
\cite{Kim:1979if,Shifman:1979if, Zhitnitsky:1980tq, Dine:1981rt}- such as
a) the appearance of a single scale $f_a$ ($f_a\sim 10^{10}-10^{12}$ GeV) which controls both their mass  and their coupling to the gauge fields, via an 
$a(x) F\tilde{F}$ operator, where $a(x)$ is the axion field and
 b) their non-thermal decoupling at the hadron phase transition, attributed to a mechanism of vacuum misalignment. The latter causes axions to be a component of cold rather than hot dark matter, even for small values of their mass, currently expected to be in the $\mu$eV-meV range.
 
The gauging of an abelian anomalous symmetry brings in a generalization of the PQ scenario. 
As extensively discussed in \cite{Coriano:2006xh,Coriano:2007fw,Coriano:2007xg,Armillis:2008vp,Coriano:2008pg,Coriano:2009zh} it enlarges the parameter space for the corresponding axion.
This construction allows to bypass the mass/coupling relation for ordinary PQ axions, which has been often softened in various analyses of "axion-like particles" \cite{Roncadelli:2017idg}. 

Original analyses of St\"uckelberg models, motivated within the theory of intersecting branes, where anomalous $U(1)$'s are present, have resulted in the identification of a special pseudoscalar field, the St\"uckelberg field $b$. 
Its mixing with the CP-odd scalar sector allows to extract one gauge invariant component, called the axi-Higgs $\chi$, whose mass and couplings to the gauge fields are model dependent. 
If string theory via its numerous possible geometric (and otherwise) compactifications \cite{Hui:2016ltb} 
provides a natural arena where axion type of fields are ubiquitously present, then the possibility that an ultralight axion of this type is a component of dark matter is quite feasible. 
As we have discussed, its ultralight nature is a natural consequence of the implementation of the construction reflecting the low energy structure of the heterotic string theory
by involving two scales, the Planck and the GUT scale. Given the mass of such axion, it is obvious that its search has to be inferred indirectly by astrophysical observations. 

In short, we have seen that St\"uckelberg models with an axion provide a new perspective on an old problem and 
allow to open up new directions in the search for the constituents of dark matter of our universe.



\end{document}